%% file: DGrupe_astro_ph.tex
\documentclass[preprint2]{aastex}

\usepackage{epsf}
\usepackage{graphics}

\newcommand{\kms}{km s$^{-1}$}
\newcommand{\cts}{cts s$^{-1}$}
\newcommand{\ax}{$\alpha_{\rm X}$}

\newcommand{\rb}[1]{\raisebox{1.5ex}[-1.5ex]{#1}}

\newcommand{\pl}{$\pm$}

\shorttitle{Soft X-ray AGN: I. The data}
\shortauthors{Grupe et al.}

\begin{document}

\input DGrupe_clipfig.tex
\useunitmm

\def \charthoffset {\hspace{0.2cm}} \def \charthsep {\hspace{0.3cm}}
\def \chartvsepcap {\vspace{0.3cm}}
\def \chartvsep {\vspace{0.1cm}}
\newcommand{\putchartb}[1]{\clipfig{#1}{75}{20}{7}{275}{192}}
\newcommand{\putchartc}[1]{\clipfig{#1}{55}{33}{19}{275}{195}}
\newcommand{\chartlineb}[2]{\parbox[t]{18cm}{\noindent\charthoffset\putchartb{#1}\charthsep\putchartb{#2}\chartvsep}}

\newcommand{\chartlinec}[3]{\parbox[t]{18cm}{\noindent\charthoffset\putchartc{#1}\charthsep\putchartc{#2}\chartvsep\putchartc{#3}\chartvsep}}


\title{A Complete Sample of Soft X-ray Selected AGN: 
I. The Data\thanks{Based in part on observations at the European Southern
Observatory La Silla (Chile) with the 2.2m telescope of the Max-Planck-Society
during MPI and ESO time, and the ESO 1.52m telescope during ESO time in September
1995 and September 1999.}
}


\author{Dirk Grupe\thanks{Guest observer, McDonald Observatory,
University of Texas at Austin} \altaffilmark{ , 3}}
\affil{Astronomy Department, Ohio State University,
    140 W. 18th Ave., Columbus, OH-43210, U.S.A.}
\email{dgrupe@astronomy.ohio-state.edu}

\author{Beverley J. Wills\altaffilmark{4}}
\affil{Astronomy Department, University of Texas at Austin, RLM 15.308, Austin,
TX 78712, U.S.A.}
\email{bev@pan.as.utexas.edu}

\author{Karen M. Leighly\altaffilmark{5}}
\affil{Dept. of Physics and Astronomy, University of Oklahoma, 440 W. Brooks
St., Norman, OK 73019, U.S.A.}
\email{leighly@nhn.ou.edu}

\author{Helmut Meusinger\altaffilmark{6}}
\affil{ Th\"uringer Landessternwarte Tautenburg, Sternwarte 5, D-07778 Tautenburg,
Germany}




\begin{abstract}
We present the optical spectra and simple statistical analysis 
for a complete sample of 110 soft X-ray selected AGN.
About half of the
sources are Narrow-Line Seyfert 1 galaxies (NLS1s), which 
have the steepest
X-ray spectra, strongest FeII emission and slightly weaker
[OIII]$\lambda$5007 emission than broad line Seyfert 1s (BLS1s).
 Kolmogorov Smirnov 
tests show that NLS1s and BLS1s have
clearly different distributions of the X-ray spectral slope \ax~, X-ray
short-term variability, and FeII equivalent widths and luminosity and
FeII/H$\beta$ ratios. The differences in the [OIII]/H$\beta$ and [OIII]
equivalent widths are only marginal. 
We found no significant differences
between NLS1s and BLS1s in their rest frame 0.2-2.0 X-ray luminosities,
rest frame 5100\AA~monochromatic luminosities, bolometric luminosities,
 redshifts,  and their H$\beta$ equivalent widths. 

{\bf Please note:} this is a special version for astro-ph that does not contain
the optical and FeII subtracted spectra. The complete paper including the
spectra can be retrievd from
http://www.astronomy.ohio-state.edu/$\sim$dgrupe/research/sample\_paper1.html
\end{abstract}


\section{Introduction}

With the launch of the X-ray satellite ROSAT (\citet{tru83}) a new chapter in
the history of astronomy was written. With the spectral sensitivity of the
Position Sensitive Proportional Counter (PSPC, \citet{pfe86}) to energies as low
as 0.1 keV it was possible for the first time to study  the soft
X-ray properties of a large number of AGN. 
In the first half year of its mission ROSAT performed, for the first time,
an all-sky survey (RASS,
\citet{vog98}) in the 0.1-2.4 keV energy band. 
 This survey led to the discovery of a large number of previously
unknown soft X-ray sources (\citet{tho98, beu99, sch00}), about 1/3 of them AGN. 
Many AGN show a strong excess in
soft X-rays. Most of their
bolometric luminosity is emitted in the energy range between the UV and soft
X-ray energies. It is commonly believed that this 'Big Blue Bump' emission is
produced by an accretion disk surrounding the central black hole
(e.g. \citet{shi78, mal82, mal83, band89}).
The soft X-ray emission can be
explained by Compton scattering of thermal UV photons in a
layer of hot electrons above the disk (e.g. \citet{czerny87, laor89, ros92,
mann95}).
The closer to the Eddington limit
the black hole accretes, the softer the X-ray spectrum is expected to become 
(e.g. \citet{ros92, pound95}). 
Alternatively, the soft X-ray emission may also be result in an optically thick
wind from the black hole region (\citet{king03}).

In the days
before ROSAT the study of strong
soft X-ray AGN depended on serendipitous
observations, e.g. by EINSTEIN (\citet{cor92, puc92}), and observations of AGN
selected at optical wavelengths.
\citet{ste89} noticed in a sample of EINSTEIN-selected AGN that more than 25\%
of her sources belonged to the Seyfert 1 (sub)class  of Narrow-Line Seyfert
galaxies (NLS1s, \citet{ost85}), while in optically selected samples 
only about
10\% of the sources are NLS1s (\citet{ost85, ost87, wil03})
This higher fraction of NLS1s among X-ray selected sources
was confirmed by \citet{puc92} for a sample
of 52 EINSTEIN-detected AGN: 9 of their 17 Seyfert 1
galaxies were NLS1s. \citet{gru96, gru99b}, and \citet{edel99} 
found that in soft X-ray selected ROSAT AGN samples
up to 40\% were NLS1s. NLS1s show extreme
properties, such as steep X-ray spectra (e.g. \citet{bol96, gru96, wil03}), 
strong optical
FeII and weak emission from the Narrow-Line Region (e.g. \citet{bor92, bor02,
gru96, lao97, gru99b}).

We have studied the continuum and emission line properties of a sample of 76
soft X-ray selected ROSAT AGN (\citet{gru96, gru98, gru99b}). 
 However, that sample was incomplete lacking a significant number of sources for
 which optical spectra were not obtained at that time.
Our new sample containing 110 sources is complete following the criteria in
\S\,\ref{sample}.  For each source 
X-ray and optical spectra exist that have enough quality to allow for a detailed
analysis of their X-ray and optical properties.
We performed a detailed study of the X-ray properties of the complete
soft X-ray selected AGN  sample
(\citet{gru01}).
Here we describe the sample selection (\S\,\ref{sample}), the observations
(\S\,\ref{observe}), and data reduction (\S\,\ref{reduction}). 
The FeII subtraction and the line measurements are described in \S\,\ref{lines}.
We present an analysis of the distributions of continuum and emission line
properties of NLS1s and BLS1s
in \S\,\ref{results}. The results will be discussed in \S\,\ref{discuss}.
Previously unpublished optical spectra are presented at the end of the paper.
In a second paper (\citet{gru03a}, Paper II)
we will present direct correlations and a Principal Component
Analysis. 

Throughout the paper spectral slopes are defined as energy spectral slopes with
$F_{\nu} \propto \nu^{-\alpha}$. Luminosities are calculated assuming a Hubble
constant of $H_0$ =75 \kms Mpc$^{-1}$ and a deceleration parameter of $q_0$ =
0.0.

\section{\label{sample} Sample Selection}

Our AGN sample was selected from the bright soft X-ray sample presented by
\citet{tho98},
using the following criteria:

\begin{itemize}
\item Mean RASS PSPC count rate $\geq$0.5 \cts
\item Hardness ratio\footnote{Hardness ratio = (hard-soft)/(hard+soft) 
with the soft energies = 0.1-0.4 keV and
hard energies = 0.5-2.0 keV.} $<$ 0.00
\item Galactic latitude $\rm |b|$ $>~20^{\circ}$
\end{itemize}

A count rate threshold of 0.5 was chosen to ensure sufficient X-ray photons during a
typical RASS time coverage of about 200-400s
to perform spectral analysis. The hardness
ratio criterion ensures a soft X-ray spectrum and the galactic latitude
criterion ensures no hardening of the X-ray spectrum due to extinction.
Using these criteria, \citet{tho98} found
397 sources of which 113 turned out to be AGN
(\citet{tho98, gru01}),
excluding BL Lac objects that have different emission mechanisms. 
 We have also excluded the three known transient sources from the present
sample (IC 3599, \citet{bra95, gru95a}; WPVS007, \citet{gru95b}, and 
RX J1624.9+7554, \citet{gru99a}), because these sources were observed to be
bright in X-rays only once
and at least in IC 3599 and RX J1624.9+7554 the X-ray emission
might be due to a dramatic accretion event (e.g. \citet{gez03}).
The nature of the X-ray transience in WPVS 007 is still unclear.. 

\section{\label{observe} Observations}

\subsection{X-ray data}
In addition to the RASS data, available for all sources, 
about 50 have pointed PSPC observations and are available from the ROSAT public
archive at MPE Garching. A detailed description of the X-ray observations and
analysis is given in \citet{gru01}.

\subsection{Optical spectroscopy}
Optical spectroscopy data in the rest frame H$\beta$ region
were collected over a period of 10 years using various
observatories and telescopes. Table \ref{obs_sum} summarizes the 
observations. The table contains the coordinates of the X-ray position of the
source in Equinox J2000, a common name, the observing date, the telescope and
instrument, the observation time and comments. The comments list 
references to spectra that have been already published and weather
conditions.
In the following we describe the
observations by telescope as they appear in Table
\ref{obs_sum}.

\subsubsection{ESO 2.2m and 1.52 m telescopes}
All observations of objects with southern declination prior to 1995 were observed
with the MPI/ESO 2.2m telescope at La Silla (ESO2.2 in Table \ref{obs_sum}). 
At that time, the 2.2m was equipped
with the ESO Faint Object Camera and Spectrograph (EFOSC), 
which had a selection
of different grisms for spectroscopy. Table \ref{obs_sum} lists the number of
the grism(s) used. The grisms had the following resolutions and wavelength
coverages:

\begin{itemize}
\item \#4: 2.2 \AA/pix $\approx$ 7 \AA~FWHM resolution, \\
4650 -- 6800 \AA
\item \#8: 1.3 \AA/pix $\approx$ 4 \AA~FWHM resolution, \\
4640 -- 5950 \AA
\item \#9: 1.1 \AA/pix $\approx$ 4 \AA~FWHM resolution, \\
5875 -- 7020 \AA
\item \#10: 1.2 \AA/pix $\approx$ 4 \AA~FWHM resolution, \\
6600 -- 7820 \AA
\end{itemize}

Slit widths were usually $1.^{''}5$ or 2$^{''}$ and the 
slit orientation was always in E-W direction.

The observations  in 1995 and 1999 were performed with
ESO's 1.52m telescope equipped with the Boller \& Chivens 
spectrograph (B\&C). At both
times, the grating \#23 with 600 grooves mm$^{-1}$ resulting in a dispersion 
in first order of 126 \AA~mm$^{-1}$ or 1.89 \AA~ Pixel$^{-1}$. The 
resolution was about 6\AA~ FWHM. For all the spectra we used a slit width of
2$^{''}$.1. All observations with the ESO 1.52m telescope were performed at
parallactic angle.

\subsubsection{McDonald Observatory 2.1m and 2.7m telescopes}

The telescope used for most northern hemisphere 
observations was the 2.1m Otto Struve telescope at McDonald Observatory.
The ES2 spectrograph, which has a  similar design to the
Boller \& Chivens spectrograph, used
grating \#22 during the 1994 run, giving 
a dispersion of
112\AA~mm$^{-1}$ ($\approx$4\AA~ FWHM resolution).
During the 1995 to 1999 runs, grating \#4 with 222\AA~mm$^{-1}$
($\approx$8\AA~ FWHM resolution) was used.  
The slit widths for the 1995 to 1999 observing runs were either 1$^{''}$.6 or
2$^{''}$.0 and the slit orientation was in E-W direction for all observations.

A number of sources were observed with the  Large Cassegrain Spectrograph (LCS)
on the 2.7m Harlan-Smith Telescope at McDonald Observatory.
Several of these sources are from an
overlapping program of studies of PG quasars (\citet{wills00, sha03}). 
The slit width of the PG quasars 
was 1$^{''}$ and 2$^{''}$ for all other objects observed with the 2.7m telescope.
The slit orientation was in E-W direction. 
The instrumental resolution was $\approx$7\AA (\citet{sha03}) for the PG quasars and
14 for the other sources. One of the PG quasars, PG 1626+554, was observed with the
Hubble Space Telescope (HST) using the Faint Object Spectrograph (FOS) with a
0.86$^{''}$ circular width.

\subsubsection{CTIO and Tautenburg}
A few objects were observed with the 4m Blanco telescope at the
Cerro Tololo International Observatory in Chile
(CTIO4.0 in Table\,\ref{obs_sum}). 
The instrument used was
the R-C spectrograph with the KPGL3 grating having a dispersion of
116 \AA~mm$^{-1}$ (4.3\AA~FWHM resolution) with a slit width of 2$^{''}$.
The slit was oriented in E-W direction.

Four sources were observed with the 2.0m telescope of the 
Th\"uringer Landessternwarte (TLS) Tautenburg, Germany,
(TLS2.0 in Table \ref{obs_sum}) during the test stage of the 
Nasmyth Focal Reducer Spectrograph (NFRS) which was constructed and built 
at TLS. The V200 grism, with 300 grooves mm$^{-1}$ 
corresponding to a dispersion of 225 \AA~ mm$^{-1}$ or 3.38 \AA~ px$^{-1}$, was
used.
The grism provides a wavelength coverage from 4\,500 to 9\,000 \AA. 
The slit width of 1\arcsec~ for B2\,1128+31 and 2\arcsec for the other 
objects yields a resolution of 7\AA~ or 14\AA, respectively. 
Slit orientation was fixed (N-S direction).

\section{\label{reduction} Data reduction}
All spectra, except for the ones taken at Tautenburg (see below), were bias and
flat-field corrected and were wavelength- and flux-calibrated by taking spectra
of calibration lamps and flux calibration standard stars. All data 
were reduced, except for the McDonald 2.7m and CTIO data, with ESO's MIDAS data
reduction package.  The McDonald 2.7m and CTIO data were
reduced using IRAF. The 1D-spectra were extracted from the two-dimensional
spectra using the optimal extraction algorithm as described by \citet{hor86}.

The data taken at TLS Tautenburg required
special treatment because no calibration files were obtained. 
The bias was
estimated from unexposed parts of the CCD. No flatfield correction was applied.

The TLS Tautenburg is located about 10 km north-east of
the city of Jena. Usually, light pollution from cities can severely harm
astronomical observations. However, in our case the illumination of the sky by
Jena's street lamps gave us a perfect night sky wavelength calibration spectrum.
\citet{ost92} describe  how the light pollution at Lick Observatory
can be used for wavelength calibration. We identified the emission lines of the
street lamps of Jena in the observed spectra and used these for the wavelength
calibration.

The flux calibration was more challenging, because no standard star was
observed. However, by chance a star was in the slit of three exposures at
$\alpha_{2000}$= 10h 40m 01.0s, $\delta_{2000}~=~21^{\circ} 08^{'} 34^{''}$.
From the colors derived from the Automatic Plate Measuring (APM) scans of the
Palomar Observatory Sky Survey plates (POSS) 
and 2 Micron All-Sky Survey (2MASS) we could identify the star as a
K3V star. 
The known spectral shape of a K3V star allowed us to correct for the
detector/telescope response and atmospheric transmission.
The flux density was then determined from the APM magnitudes of the K3V
star, using the K3V star RX J1320.7+0701 (\citet{tho98}) as a reference.
 
\begin{figure}
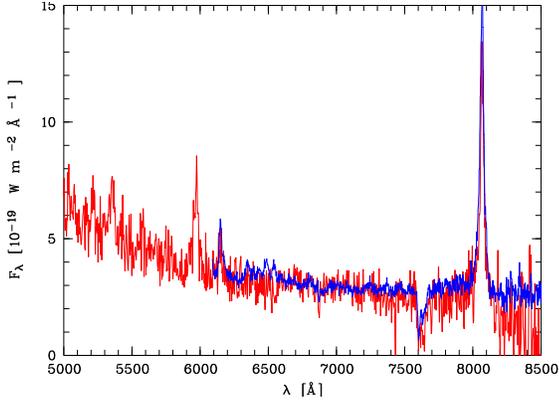

\clipfig{DGrupe.fig1}{75}{20}{7}{280}{192}
\caption{\label{rxj1304_mcd_tls} Combined spectrum of spectra of RX J1304.2+0205
taken at McDonald Observatory (black) and the Th\"uringer Landessternwarte
Tautenburg (grey)
}
\end{figure}

Figure\,\ref{rxj1304_mcd_tls} shows the spectrum of RX J1304.2+0205 taken with the
2.7m at McDonald Observatory and the TLS Tautenburg.
The figure demonstrates how well the calibration of the Tautenburg data matches
the McDonald Observatory data.

\section{\label{lines} Line measurements}

\subsection{FeII subtraction}
In general,
Seyfert 1 galaxies show FeII emission in their optical
spectra. The strength of the FeII emission is correlated with the softness of
the X-ray 
spectrum and anti-correlated with the strength of the [OIII] emission (e.g.
\citet{bor92, bor02, gru99b}. In order to accurately measure the [OIII] and
H$\beta$ emission lines, the FeII emission must be removed from the spectrum.
This is especially important for sources with weak [OIII] emission, which
are the ones with the strongest FeII emission.

We adopt the
method described by \citet{bor92}, using the FeII template
of I\,Zw\,1 given in their paper for the wavelength range $\sim$
4400-6000\AA.  In order to correct also for the bluer part of the
spectrum, this template was extended towards shorter wavelengths 
using the relative FeII line intensities given by \citet{phi78a,phi78b}.
The whole template was wavelength-shifted according to the
redshift of the object's spectrum and the individual FeII lines were
broadened to the FWHM of the broad H$\beta$ line by using a Gaussian
filter.  The template was scaled by eye to match the line intensities of the
object spectrum and then subtracted. 

The FeII rest frame equivalent width  and flux were measured in the
rest frame range between 4430-4700\AA ($\lambda$4570\AA~blend)
from the
redshifted and scaled template. 
This wavelength range
allows a direct comparison with the EW(FeII) given in \citet{bor92}.
By measuring the FeII flux and equivalent width from the template
instead of the source spectrum, we avoid contamination of the
HeII$\lambda$4686 line to the measurements. Note that the FeII/H$\beta$ and
EW(FeII) values given for the old sample (\citet{gru99b}) were
based on the flux in
the entire template between 4250\AA-5880\AA.

The method of subtracting an FeII template of the NLS1 I Zw 1 from the AGN
spectrum works well for most of the sources (e.g. RX J2242.6--3845 or
MS 23409--1511).
However, in some cases the FeII subtracted spectrum shows large
residuals. Measurements of the FeII equivalent widths and fluxes 
from these sources have to be taken with caution.  
This is the case for the following objects: Mkn 335, Mkn 1044, RX J1005.7+4332,
Mkn 141, Mkn 142, NGC 4051, QSO 1421--0013, and NGC 7214.

The reason for these discrepancies is that the template used was derived from
one galaxy (I Zw 1) which is likely to have different physical
 conditions (temperature,
electron density, ionization spectrum). The strengths of the different FeII
atomic transitions  depend strongly on these conditions 
(e.g. \citet{sig03} and \citet{ver03}). If
the conditions in the BLR of 
our sources deviate from those in I Zw 1, we can
expect that the template will not completely work to subtract and describe the
FeII emission in this source.

\subsection{Emission line and continuum parameters}

Table \ref{line_sum} summarizes the redshifts, the FWHM, rest frame 
equivalent widths (EW), and the [OIII]/H$\beta$ and FeII/H$\beta$
line ratios
measured from the optical spectra. 
The FWHM of the H$\beta$ and [OIII]$\lambda$5007 lines were
measured directly from the lines. 
For the H$\beta$ line, FWHM, EW and flux were measured in the broad component 
after subtracting a narrow
component. This narrow component was constructed from an 
appropriately scaled template of the
[OIII]$\lambda$5007 line (\citet{gru98b,gru99b}).
All FWHM are given in the rest frame and are corrected for instrumental
broadening assuming that
$FWHM_{\rm true}=\sqrt{FWHM^2_{\rm obs}-FWHM^2_{\rm instr}}$.
Due to
the differing weather conditions for the observations, no line fluxes are given,
only line ratios.
Note  that the [OIII]/H$\beta$ line ratio is the flux of the
[OIII]$\lambda$5007 line to the broad component of H$\beta$  and not
the narrow component, as used in the diagrams of \citet{vei87}. 

Errors in the FWHM are typically
$\approx$ 10\% for the H$\beta$ line,
 but are larger for the  [OIII] line due to the contamination of FeII
and uncertainties in corrections for observational and instrumental resolution.
The errors in the equivalent widths
are critically dependent on determination of the continuum and can be as much as
25\%, and in some cases where the FeII subtraction was less satisfactory they
can be even larger. Another source of uncertainty
in the equivalent width is in the contribution
of star light from the host galaxy. This is more important for the
low-luminosity than for the high-luminosity AGN. Additional errors of the H$\beta$
FWHM and EW are the uncertainties of the subtraction of a narrow H$\beta$ component.
While it is quite easy to subtract a narrow H$\beta$ component in Seyfert 1.5
galaxies, such as Mkn 1048, it is more complicated in NLS1s. Adjusting the
[OIII]$\lambda$ 5007 template by eye usually results in a [OIII]/H$\beta_{\rm narrow}$
ratio of about 3 while this ratio is typically 10-15 in Seyfert 1.5s (e.g.
\citet{cohen83}).
We therefore
measured the H$\beta$ line twice, once with an [OIII]/H$\beta_{\rm narrow}$ ratio
determined from the template adjusted by eye and second from a [OIII]/H$\beta_{\rm
narrow}$=10 ratio of the template.
This was used to estimate the error in the H$\beta$ FWHM and EW.
The optical continuum
and line luminosities for an individual source can be in error
by a factor of two to
three due to non-photometric weather, bad seeing, and slit losses.

\begin{figure}
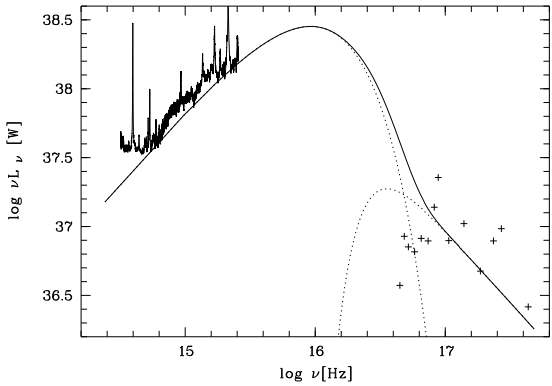

\clipfig{DGrupe.fig2}{75}{15}{7}{280}{192}
\caption{\label{rxj1118_sed} Powerlaw with exponential cutoff and powerlaw with
neutral absorption (solid line)
to the optical/UV (\citet{sha03}) and X-ray spectrum of PG
1115+407 as an example how the bolometric luminosity was determined. The 
dotted lines display the powerlaw with exponential cutoff for the optical/UV
part and the powerlaw with neutral absorption for the soft X-ray part
separately.
}
\end{figure}

Table \ref{lum_sum} lists the soft X-ray 0.2-2.0 keV
spectral index \ax~(\citet{gru01}),
 the rest frame 0.2-2.0 X-ray luminosity $L_{\rm X}$, 
the optical monochromatic luminosity at
5100\AA, $\lambda L_{5100}$,  the bolometric luminosity $L_{\rm bol}$, and the
soft X-ray short-term variability parameter $\chi^2/\nu$ (\citet{gru01}). 
The bolometric luminosity was estimated from a combined
powerlaw model fit with exponential cutoff to the optical-UV data and a powerlaw
with absorption due to neutral elements,
to the soft X-ray data (see Figure\,\ref{rxj1118_sed}). 
Note that because the EUV part of the spectral energy
distribution of AGN is unobservable, there are large uncertainties in the
bolometric luminosity (e.g. \citet{elv94})  
and the values given here are only approximate.

\section{\label{results} Results}

We classified all objects with FWHM(H$\beta$)$\leq$2000 \kms~as Narrow Line
Seyfert 1s (NLS1s) and all sources with FWHM(H$\beta$)$>$2000 \kms~as Broad Line
Seyfert 1s (BLS1s) following the definition of \citet{ost85} and \citet{good89}
regardless of subgroups such as Sy\,1.5s. This results in 51 NLS1s and 59 BLS1s.

\subsection{Simple Statistics}

Table\,\ref{distr_sum} summarizes the mean, standard deviations, medians of the
FWHM of H$\beta$ and [OIII], the rest frame
equivalent widths of H$\beta$, [OIII] and FeII,
the [OIII]/H$\beta$ and FeII/H$\beta$ flux ratios,
the X-ray slopes \ax, the rest frame 0.2-2.0 keV X-ray luminosities, the 5100\AA~
rest frame monochromatic luminosities, the bolometric luminosities, 
and the redshifts  of the whole sample of 110
AGN, the 51 NLS1s, and the 59 BLS1s. 
NLS1s have in general the steepest X-ray spectra and
strongest FeII emission, shown both in the equivalent width of FeII as well as
in the FeII/H$\beta$ line ratio. 
There are no differences between NLS1s and BLS1s with respect to their mean 
equivalent width of H$\beta$ and luminosities in X-rays and at 5100\AA.
Kolmogorov-Smirnov KS tests for these properties show that the 
distributions are similar.

\begin{figure*}
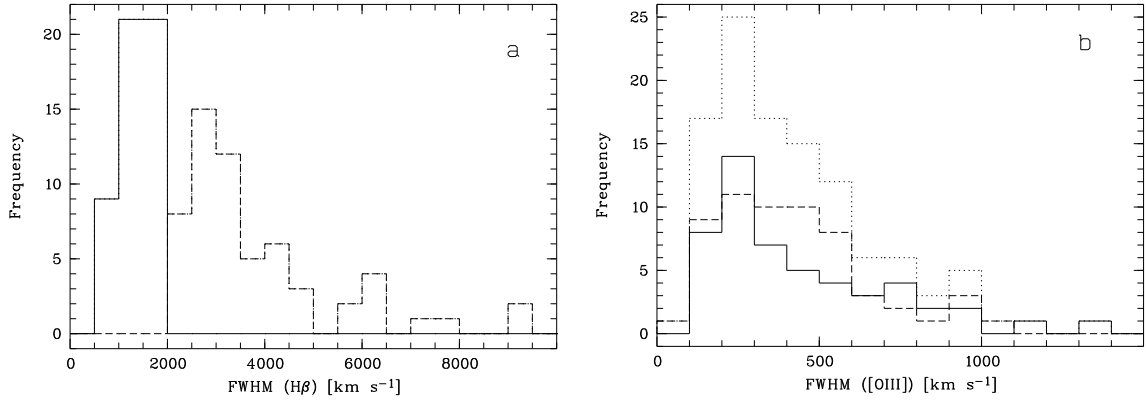

\chartlineb{DGrupe.fig3a}{DGrupe.fig3b}
\caption{\label{distr_fwhm} Distribution of the FWHM of
H$\beta$ and [OIII] for the whole sample (dotted line), NLS1s (solid
line), and BLS1s (short dashed line). FWHM have been corrected for instrumental
resolution.
}
\end{figure*}

\begin{figure*}
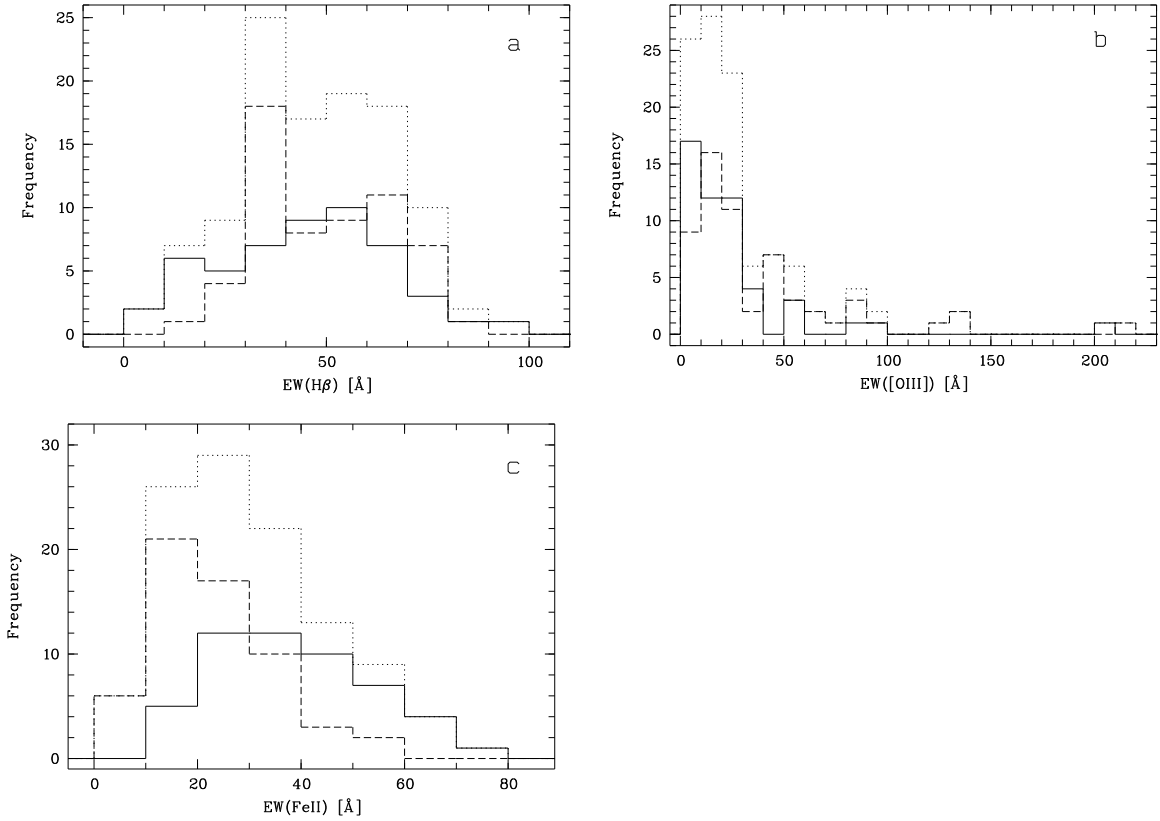

\chartlineb{DGrupe.fig4a}{DGrupe.fig4b}
\clipfig{DGrupe.fig4c}{75}{20}{7}{275}{192}
\caption{\label{distr_ew} Distribution of the rest frame
Equivalent widths of
H$\beta$, [OIII], and FeII. 
The lines are the same as described in Figure\,\ref{distr_fwhm}. 
}
\end{figure*}

Figure\,\ref{distr_fwhm} displays
the distributions of the FWHM of H$\beta$ and [OIII]. 
Due to their definition, they have different distributions of their
FWHM(H$\beta$).
NLS1s and BLS1s have similar distribution in their
 FWHM([OIII]).
Figure\,\ref{distr_ew} shows the distributions of the equivalent widths of
H$\beta$, [OIII], and FeII. There are no differences in the
distributions of the EW(H$\beta$) (Figure \ref{distr_ew}a), but the EW(FeII) distributions 
(Figure\,\ref{distr_ew}c) are different at a level $P>$99.99\%.
NLS1s dominate the group with small values
of the EW([OIII]).  
KS tests indicate that 
the EW([OIII]) distributions (Figure\,\ref{distr_ew}b) of NLS1s and BLS1s
are only marginally different with a probability of 99.1\%

\begin{figure}
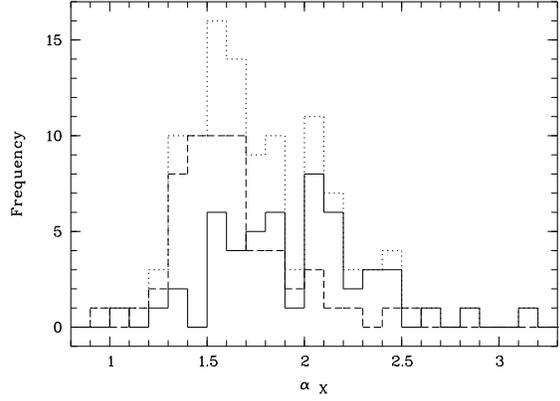

\clipfig{DGrupe.fig5}{75}{20}{7}{275}{192}
\caption{\label{distr_ax} Distribution of the X-ray spectral slope \ax.
The lines are the same as described in Figure\,\ref{distr_fwhm}.
}
\end{figure}

Figure\,\ref{distr_ax} displays the distributions of the X-ray spectral index in the
ROSAT PSPC energy range. 
NLS1s have, as expected, the steepest X-ray spectra while BLS1s show
flatter X-ray spectra. There is a $>$99.99\% probability that the
distributions are different.

\begin{figure*}
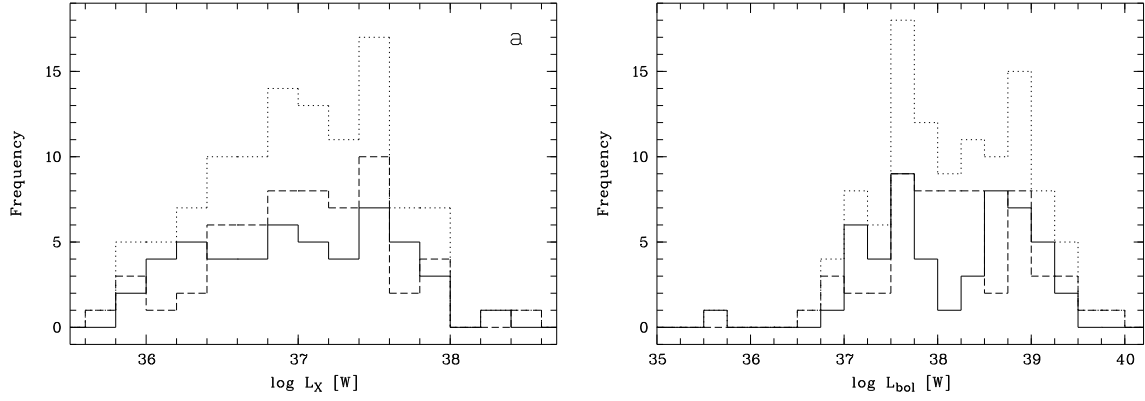

\chartlineb{DGrupe.fig6a}{DGrupe.fig6b}
\caption{\label{distr_xray} Distribution of
the rest frame 0.2-2.0 keV  X-ray luminosity $L_{\rm X}$ and the bolometric
luminosity $L_{\rm bol}$.
The lines are  as described in Figure\,\ref{distr_fwhm}.
}
\end{figure*}

\begin{figure}
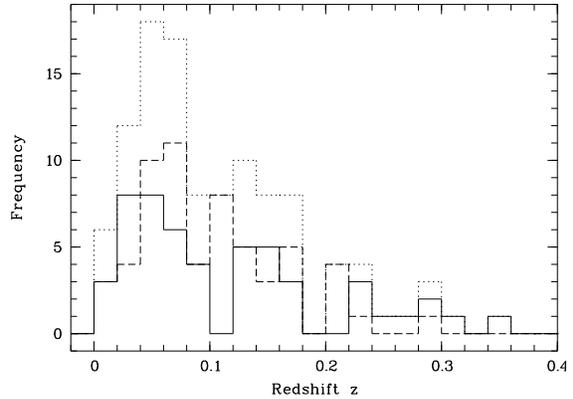

\clipfig{DGrupe.fig7}{75}{20}{7}{275}{192}
\caption{\label{distr_z} Distribution of the redshift z.
The lines are the same as described in Figure\,\ref{distr_fwhm}.
}
\end{figure}

The distributions in the rest frame 0.2-2.0 X-ray luminosities and the
bolometric luminosities 
 are shown in Figure\,\ref{distr_xray}. There are no
significant differences in any of the luminosity distributions between
  NLS1s and BLS1s. 
Figure\,\ref{distr_z}
displays the redshift distributions of the two samples. 
A KS-test shows that the distributions are similar.

\begin{figure*}
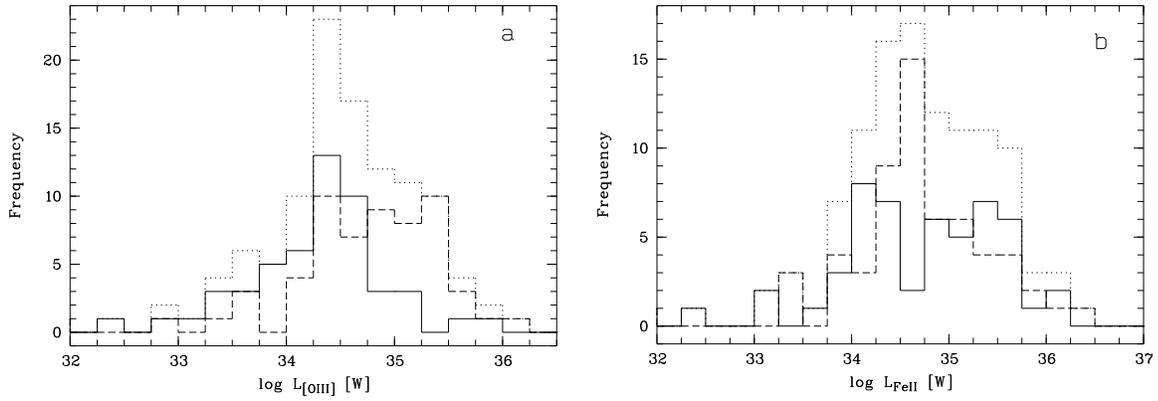

\chartlineb{DGrupe.fig8a}{DGrupe.fig8b}
\caption{\label{distr_line_lum} Distribution of the [OIII] (left) and
FeII luminosities (right).
The lines are the same as described in Figure\,\ref{distr_fwhm}.
}
\end{figure*}

\begin{figure*}
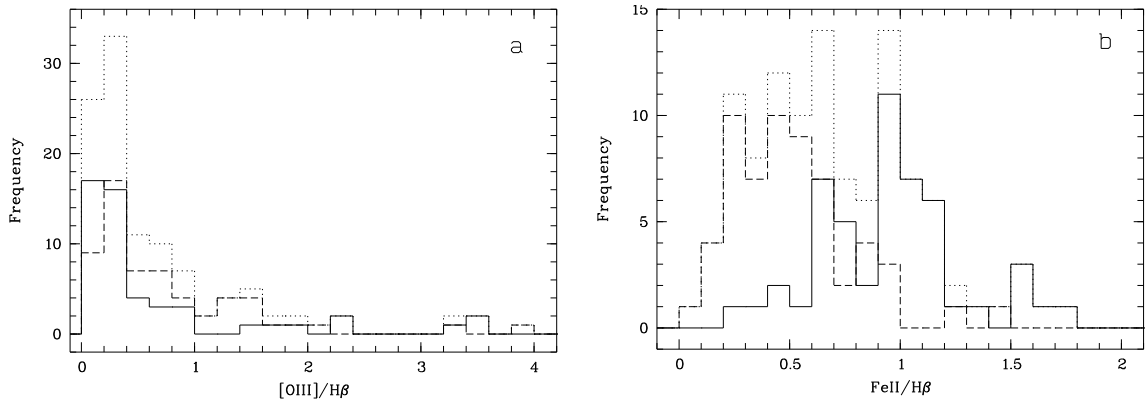

\chartlineb{DGrupe.fig9a}{DGrupe.fig9b}
\caption{\label{distr_ratios} Distribution of the [OIII]/H$\beta$ and
FeII/H$\beta$ flux ratios.
The lines are the same as described in Figure\,\ref{distr_fwhm}.
}
\end{figure*}

Figure\,\ref{distr_line_lum} displays the distributions of the [OIII] and FeII
luminosities. A KS test shows that NLS1s and BLS1s have
different distributions in their FeII luminosity ($P>$99.99\%), but show similar
distributions in their [OIII] luminosity. 
Figure\,\ref{distr_ratios} shows the distributions of the 
[OIII]/H$\beta$ and FeII/H$\beta$ flux ratios. While a KS test shows that the
distributions of FeII/H$\beta$ of NLS1s and BLS1s are clearly different, there
is a 3\% chance that the [OIII]/H$\beta$ distributions  are similar.  

\begin{figure}
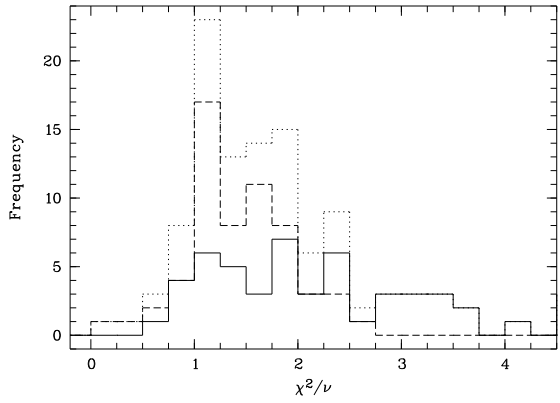

\clipfig{DGrupe.fig10}{75}{20}{7}{275}{192}
\caption{\label{distr_chisqu} Distribution of the X-ray variability parameter 
$\chi^2/\nu$. Outside the plot are NGC 4051, Mkn 766, and RX J1304.2+0205 with
$\chi^2/\nu$=18.3, 8.86, and 5.08, respectively.
The lines are the same as described in Figure\,\ref{distr_fwhm}.
}
\end{figure}

Figure\,\ref{distr_chisqu} shows the distributions of the short-term variability
parameter $\chi^2/\nu$ of the RASS observations (\citet{gru01}). A KS test shows
that the distributions are different with a probability of P$>$99.99\%.

\subsection{Spectra}

Figure\,\ref{opt_spectra} displays all optical spectra in the observed frame
that have not been published
yet. For many of the
sources, optical spectra are shown here for the first time. In
other cases, spectra have been published before as listed in Table \ref{obs_sum},
but the spectra shown here are either of better quality or show a wider
wavelength coverage. The spectra previously published by \citet{gru99b} can be
accesses electronically at the CDS anonymous FTP site at {\it ftp 130.79.128.5}.

Figure\,\ref{fe2_spectra} displays the FeII-subtracted spectra, the template used
and the original spectra before template subtraction.
The FeII subtracted spectra are shown
with their calibrated fluxes. The FeII template and the original spectra are
offset.

{\bf Please note:} this is a special version for astro-ph that does not contain
the optical and FeII subtracted spectra due to disk space limitations.
The complete paper including the
spectra can be retrievd from
http://www.astronomy.ohio-state.edu/$\sim$dgrupe/research/sample\_paper1.html

\section{\label{discuss} Discussion}

\subsection{Statistics}
The main aims of this paper are to present the spectra and a simple statistical
analysis of our sample of 110 soft X-ray selected AGN. 
About half of the sources of our complete sample of 
soft X-ray selected AGN turn out
to be NLS1s (51). 
In a second paper (\citet{gru03a}) 
a Principal Component Analysis (PCA) will be shown. 
The NLS1s show the well-known
\citet{bor92} Eigenvector 1 relation between [OIII] and FeII. 
It is not a complete surprise 
that about half of our sources are NLS1s, because NLS1s do have the
steepest X-ray spectra among AGN 
(e.g. \citet{bol96, gru99b, gru01, gru00}) and
the best way to find them is through a soft X-ray survey.

It is surprising that there is no significant
difference between the
rest frame 0.2-2.0 keV X-ray and 5100\AA, and bolometric
luminosities for NLS1s and BLS1s. 
Even though the sources are variable in X-rays (\citet{gru01}) the rest frame
0.2-2.0 keV X-ray luminosity is a better measure of 
power output of the nucleus than the optical luminosity here because
of the different weather
conditions, it is impossible to get absolute flux calibrated spectra if
the conditions were not photometric. The other reason is that most of the
spectra have been taken with 2$^{''}$ slit widths, which means that the
contribution of galactic starlight can be significant, especially for the
low-luminosity AGN. 
The contribution to the total X-ray luminosity 
of sources such as high-luminous X-ray binaries of supernova
remnants, is on the order of about
$10^{31}-10^{33}$ W, and 
is therefore negligible compared with the X-ray power of the nucleus.

The usual interpretation of the steep X-ray spectra of NLS1s is that
these sources accrete close to their Eddington limit (e.g. \citet{pound95}). 
If this is
true and the distributions in luminosity of all types of sources in our sample
are the same we can interpret this result to imply
that in general NLS1s have
smaller central black hole masses than BLS1s as suggested by
\citet{bol96} and \citet{wan98} and shown for some NLS1s by e.g. 
\citet{onken03}.
This assumption applies also for our sample. \citet{gru03b} found, that for the
soft X-ray selected sample presented here, for a given luminosity, NLS1s have
smaller black hole masses than BLS1s when the black hole masses were
determined by the relations given in \citet{kas00}. It will also be shown in
that paper, that NLS1s show a different $M_{\rm BH}~-~\sigma$ relation
(\citet{mag98, tre03}) than non-active galaxies and broad-line Seyferts.

\subsection{[OIII] and FeII strengths}

There are two interesting things about the distributions of [OIII] and FeII
strengths: a) for the luminosity distributions NLS1 and BLS1s have similar FeII
luminosity distributions but different [OIII] luminosity distributions, and b)
surprisingly the [OIII]/H$\beta$ distributions are only slightly different with a
3\% change of being similar, but the
FeII/H$\beta$ distributions are clearly
different. From the \citet{bor92} Eigenvector 1
relation, which we also see in our sample (Paper II), one would expect that also
the FeII luminosity and [OIII]/H$\beta$ distributions of NLS1s and BLS1s would
be different. On the other hand, 
NLS1s and and BLS1s do show slightly different
distributions of their [OIII] equivalent widths and clearly different
distributions of their FeII$\lambda$4560 equivalent widths.

One has to keep in mind that due to the slit widths of about
2$^{''}$, a contribution from circumnuclear HII regions cannot be excluded.
The contribution of hydrogen absorption lines 
of a medium age stellar population can also be a problem. 
The typical contribution of this component is on
the order of 2-3 \AA~ in the equivalent widths of the Balmer lines (e.g.
\citet{shi78b, mcc85, ho97, pop00, meu02}. The H$\beta$ equivalent widths of 
the sources of our sample are much larger than  this value 
(Tab\,\ref{line_sum} and \ref{distr_sum}) and the effect of stellar absorption
lines can be neglected.

\section{Conclusions}

We have presented the optical data of a complete sample of 110 soft X-ray
selected AGN. We found that

\begin{itemize}

\item About half of the sources are NLS1s based on their
FWHM(H$\beta$)$\leq$2000 \kms.
\item NLS1s and BLS1s show clearly different
 distributions of their \ax, rest frame 
equivalent widths of FeII, their FeII/H$\beta$ ratios, FeII luminosities,
and soft X-ray variability. KS tests also suggest slightly different 
distributions of
[OIII]/H$\beta$ and EW([OIII]).
\item NLS1s and BLS1s show similar distributions in their redshifts, continuum
luminosities, and equivalent widths of H$\beta$
\item The similar continuum
luminosities in both sub-samples and the high
accretion to mass ratios in NLS1s suggests that these have smaller black hole
masses for a given luminosity than BLS1s.
\end{itemize}

Correlation analysis of the sample including a Principal Component Analysis
will be presented in a second paper (\citet{gru03a}). The black hole masses and
their relation to the \citet{mag98} and \citet{tre03}
 $M_{\rm bh}~-~\sigma$ relation will be
presented in \citet{gru03b}.

\acknowledgments

We would like to thank Marianne Vestergaard 
for comments and suggestions on the
manuscript, the anonymous referee for a fast referee report that helped
improving the paper,
Hans-Christoph Thomas for providing the spectrum 
of the K3V star RX J13020.7+0701, and Brad Peterson for providing a spectrum of
NGC 4593. 
We also want to thank the night assistants and technical people at La Silla and
McDonald observatory, namely, Hector Vega, Dave Doss, Jerry Martin and Earl
Green. Without their support this project would not have been possible.
This research 
has made use of the NASA/IPAC Extra-galactic
Database (NED) which is operated by the Jet Propulsion Laboratory, Caltech,
under contract with the National Aeronautics and Space Administration. 
The ROSAT project is supported by the Bundesministerium f\"ur Bildung
und  Forschung (BMBF/DLR) and the Max-Planck-Society.

\clearpage


\begin{deluxetable}{rllllllll}
\tabletypesize{\scriptsize}
\tablecaption{Summary of the optical spectroscopy of the soft X-ray selected
sample
 \label{obs_sum}}
\tablewidth{0pt}
\tablehead{
\colhead{No.} &  \colhead{$\alpha_{2000}$} & \colhead{$\delta_{2000}$}  &
\colhead{Name} & \colhead{UT date}  &
\colhead{Tel.} & \colhead{Instrument} & \colhead{$T_{\rm exp}$} &
\colhead{comments} 
}
\startdata
1 & 00 06 19.5 & +20 12 11 & Mkn 335 & 1999/09/13 & ESO1.52 & B\&C & 30 & \\
2 & 00 25 00.2 & $-$45 29 34 & ESO 242$-$G8 & 1993/09/14 & ESO2.2 & EFOSC 1,8,10 &
5,30,30 & \citet{gru99b} \\
3 & 00 57 20.2 & $-$22 22 57 & Ton S 180 & 1992/10/17 & ESO2.2 & EFOSC 1,8,10 &
5,30,30 & \citet{gru99b} \\
4 & 00 58 37.4 & $-$36.06.05 & QSO 0056$-$36 & 1993/10/12 & ESO2.2 & EFOSC 1,8,10 &
10,25,20 & \citet{gru99b} \\
5 & 01 00 27.1 & $-$51 13 54 & RX J0100.4$-$5113 & 1993/10/12 & ESO2.2 & EFOSC
1,8,10 & 10,40,35 & \citet{gru99b} \\
6 & 01 05 38.8 & $-$14 16 14 & RX J0105.6$-$1416 & 1999/09/12 & ESO1.52 & B\&C & 30
& light clouds, 2-3$^{''}$seeing \\
7 & 01 17 30.6 & $-$38 26 30 & RX J0117.5$-$3826 & 1999/09/14 & ESO1.52 & B\&C & 40
& \\
8 & 01 19 35.7 & $-$28 21 32 & MS 0117$-$28 & 1995/10/01 & ESO1.52 & B\&C & 30 & \\
9 & 01 28 06.7 & $-$18 48 31 & RX J0128.1$-$1848 & 1999/09/12 & ESO1.52 & B\&C & 30
& seeing 2-3$^{''}$ \\
10 & 01 29 10.7 & $-$21 41 57 & IRAS F01267$-$217 & 1993/12/18 & ESO2.2 & EFOSC
1,8,10 & 5,30,20 & \citet{gru99b} \\
11 & 01 48 22.3 & $-$27 58 26 & RX J0148.3$-$2758 & 1992/08/26 & ESO2.2 & EFOSC
8,10 & 15,15 & \citet{gru99b} \\
12 & 01 52 27.1 & $-$23 19 54 & RX J0152.4$-$2319 & 1993/09/12 & ESO2.2 & EFOSC
1,8,10 & 5,30,30 & \citet{gru99b} \\
13 & 02 30 05.5 & $-$08 59 53 & Mkn 1044 & 1999/09/13 & ESO1.52 & B\&C & 30 & \\
14 & 02 34 37.8 & $-$08 47 16 & Mkn 1048 & 1999/09/13 & ESO1.52 & B\&C & 30 & light
cirrus, seeing 2.2$^{''}$ \\
15 & 03 11 18.8 & $-$20 46 19 & RX J0311.3$-$2046 & 1999/09/13 & ESO1.52 & B\&C & 45
& \\
16 & 03 19 48.7 & $-$26 27 12 &  RX J0319.8$-$2627 & 1992/10/18 & ESO2.2 & EFOSC
1,8,10 & 5,30,30 & \citet{gru99b} \\
17 & 03 23 15.8 & $-$49 31 11 & RX J0323.2$-$4931 & 1993/08/21 & ESO2.2 & EFOSC
1,8,10 & 30,30,30 & \citet{gru99b} \\
18 & 03 25 02.2 & $-$41 54 18 & ESO 301$-$G13 & 1993/09/13 & ESO2.2 & EFOSC 1,8,10 &
5,45,45 & \citet{gru99b} \\
19 & 03 33 40.2 & $-$37 06 55 & VCV 0331$-$37 & 1993/09/14 & ESO2.2 & EFOSC 1,8,10 &
5,30,30 & \citet{gru99b} \\
20 & 03 49 07.7 & $-$47 11 04 & RX J0349.1$-$4711 & 1993/10/11 & ESO2.2 & EFOSC 1,9
& 10,40 & \citet{gru99b} \\
21 & 03 51 41.7 & $-$40 28 00 & Fairall 1116 &  1993/10/12 & ESO2.2 & EFOSC 1,8,10
& 5,30,20 & \citet{gru99b} \\
 22 & 04 05 01.7 & $-$37 11 15 & ESO 359$-$G19 & 1999/09/12 & ESO1.52 & B\&C & 45 & \\
23 & 04 12 41.5 & $-$47 12 46 & RX J0412.7$-$4712 & 1993/10/19 & ESO2.2 & EFOSC
1,8,10 & 5,30,30 & \citet{gru99b} \\
24 & 04 26 00.7 & $-$57 12 02 & 1H 0419$-$577 &  1993/09/14 & ESO2.2 & EFOSC 1,8,10
& 5,20,15 & \citet{tur99} \\
25 & 04 30 40.0 & $-$53 36 56 & Fairall 303 & 1993/10/10 & ESO2.2 & EFOSC 1,8,10 &
5,30,20 & \citet{gru99b} \\
26 & 04 37 28.2 & $-$47 11 30 & RX J0437.4$-$4711 & 1992/08/26 & ESO2.2 & EFOSC 8,10 & 
15,15 & \citet{gru99b} \\
27 & 04 39 38.7 & $-$53 11 31 & RX J0439.6$-$5311 & 1999/09/14 & ESO1.52 & B\&C & 45
& \\
28 & 04 41 22.5 & $-$27 08 20 & H 0439$-$272 & 1999/09/13 & ESO1.52 & B\&C & 30 & \\
29 & 06 15 49.6 & $-$58 26 06 & 1 ES 0614$-$584 & 1999/09/12 & ESO1.52 & B\&C & 30 &
\\
30 & 08 59 02.9 & +48 46 09 & RX J0859.0+4846 & 1998/03/04 & McD2.1 & ES2 G4 &
2$\times$45 & \\
31 & 09 02 33.6 & $-$07 00 04 & RX J0902.5$-$0700 & 1999/03/20 & McD2.1 & ES2 G4 &
45 & \\
32 & 09 25 13.0 & +52 17 12 & Mkn 110 & 1995/03/07 & McD2.1 & ES2 G4 & 20 & \\
33 & 09 56 52.4 & +41 15 22 & PG 0953+414 & 1996/05/13 & McD2.7 & LCS & 35 &
\citet{wills00} \\
34 & 10 05 41.9 & +43 32 41 & RX J1005.7+4332 & 1995/04/02 & McD2.1 & ES2 G4 & 50
& light clouds \\
35 & 10 07 10.2 & +22 03 02 & RX J1007.1+2203 & 1998/02/21 & McD2.1 & ES2 G4 & 60 &
\\
36 & 10 13 03.2 & +35 51 24 & CBS 126 & 1995/03\tablenotemark{1} 
& McD2.1 & ES2 G4 & 160\tablenotemark{1} & \tablenotemark{1} \\
37 & 10 19 00.5 & +37 52 41 & HS1019+37 & 1999/03/13 & McD2.7 & LCS & 45 & \\
38 & 10 19 12.6 & +63 58 03 & Mkn 141 & 1995/03\tablenotemark{2} & McD2.1 & ES2 G4
& 110\tablenotemark{2} & \tablenotemark{2} \\
39 & 10 25 31.3 & +51 40 35 & Mkn 142 & 1995/03\tablenotemark{3} & McD2.1 & ES2 G4
& 190\tablenotemark{3} & \tablenotemark{3} \\
40 & 10 34 38.6 & +39 38 28 & RX J1034.6+3938 & 1998/03/04 & McD2.1 & ES2 G4 & 45
& \\
41 & 11 17 10.1 & +65 22 07 & RX J1117.1+6522 & 1995/03\tablenotemark{4} & McD2.1
& ES2 G4 & 280\tablenotemark{4} & \tablenotemark{4} \\
42 & 11 18 30.4 & +40 25 55 & PG 1115+407 & 1997/02/13 & McD2.7 & LCS & 45 &
\citet{wills00} \\
43 & 11 19 08.7 & +21 19 18 & Ton 1388 & 1995/03\tablenotemark{5} & McD2.1 & ES2
G4 & 155\tablenotemark{5} & \tablenotemark{5} \\
44 & 11 31 04.8 & +68 51 53 & EXO 1128+69 & 1998/03/04 & McD2.1 & ES2 G4 & 45 &
seeing 2.5$^{''}$ \\
45 & 11 31 09.5 & +31 14 06 & B2 1128+31 & 2002/02/14 & TLS2.0 & NFRS V200 &
4$\times$15 & seeing $>$ 3$^{''}$ \\
46 & 11 38 49.6 & +57 42 44 & SBS 1136+579 & 1999/03/21 & McD2.1 & ES2 G4 & 60 &
\\
47 & 11 39 13.9 & +33 55 51 & Z 1136+3412 & 1999/03/20 & McD2.1 & ES2 G4 & 45 &
\\
48 & 11 41 16.2 & +21 56 21 & Was 26 & 1998/02/20 & McD2.1 & ES2 G4 & 45 &
clouds \\
49 & 11 44 29.9 & +36 53 09 & CASG 855 & 1998/02/21 & McD2.1 & ES2 G4 & 45 & \\
50 & 12 01 14.4 & $-$03 40 41 & Mkn 1310 & 1998/03/04 & McD2.1 & ES2 G4 & 45 &
seeing 2.5$^{''}$ \\
51 & 12 03 09.5 & +44 31 50 & NGC 4051 & 1999/03/13 & McD2.7 & LCS & 15 & \\
52 & 12 04 42.1 & +27 54 12 & GQ Com & 1996/04/17 & McD2.7 & LCS & 40 &
\citet{sha03} \\
53 & 12 09 45.2 & +32 17 02 & RX J1209.8+3217 & 1999/03/14 & McD2.7 & LCS & 45 & \\
54 & 12 14 17.7 & +14 03 13 & PG 1211+143 & 1997/03/15 & McD2.1 & ES2 G4 & 20 &
clouds \\
55 & 12 18 26.6 & +29 48 46 & Mkn 766 & 1997/03/12 & McD2.1 & ES2 G4 & 30 &
clouds, \citet{gru98b}  \\
56 & 12 29 06.7 & +02 03 09 & 3C 273 & 1996/02/15 & McD2.7 & LCS & 10 &
\citet{sha03} \\
57 & 12 31 36.6 & +70 44 14 & RX J1231.6+7044 & 1999/03/22 & McD2.1 & ES2 G4 &
2$\times$30 & some clouds \\
58 & 12 32 03.6 & +20 09 30 & Mkn 771 & 1998/02/21 & McD2.1 & ES2 G4 & 45 &
seeing $>3^{''}$ \\
59 & 12 33 41.7 & +31 01 03 & CBS 150 & 1999/03/22 & McD2.1 & ES2 G4 &
2$\times$30 & light clouds \\
60 & 12 36 51.2 & +45 39 05 & MCG+08$-$23$-$067 & 1999/03/20 & McD2.1 & ES2 G4 & 45 
\\
61 & 12 39 39.4 & $-$05 20 39 & NGC 4593 & 1989/12/31 & OSU 1.8 & CCDS & 40 &
\citet{die94} \\
62 & 12 42 10.6 & +33 17 03 & IRAS F12397+3333 & 1997/03/12 & McD2.1 & ES2 G4 &
2$\times$45 & clouds, \citet{gru98b} \\
63 & 12 46 35.2 & +02 22 09 & PG 1244+026 & 1998/03/04 & McD2.1 & ES2 G4 & 45 &
seeing$>3{''}$ \\
64 & 13 04 17.0 & +02 05 37 & RX J1304.2+0205 & 2002/02/14 & TLS2.0 & NFRS V200 &
4$\times$15 & \\
65 & 13 09 47.0 & +08 19 48 & PG 1307+085 & 1999/03/21 & McD2.1 & ES2 G4 & 30 &
\\
66 & 13 14 22.7 & +34 29 39 & RX J1314.3+3429 & 1995/03\tablenotemark{6} &
McD2.1 & ES2 G4 & 310\tablenotemark{6} & \tablenotemark{6} \\
67 & 13 19 57.1 & +52 35 33 & RX J1319.9+5235 & 1999/03/13 & McD2.7 & LCS  &
45 & \\
68 & 13 23 49.5 & +65 41 48 & PG 1322+659 & 1996/04/18 & McD2.7 & LCS & 40 &
\citet{sha03} \\
69 & 13 37 18.7 & +24 23 03 & IRAS 13349+2438 & 1987/03/31 & McD2.7 & 
IDS\tablenotemark{7} & N/A &  \citet{wil92} \\
70 & 13 43 56.7 & +25 38 48 & Ton 730 & 1999/03/20 & McD2.1 & ES2 G4 & 45 & \\
71 & 13 55 16.6 & +56 12 45 & RX J1355.2+5612 & 1995/04/01 & McD2.1 & ES2 G4 &
45 & \\
72 & 14 05 16.2 & +25 55 34 & PG 1402+261 & 1996/05/10 & McD2.7 & LCS & 35 &
\citet{wills00} \\
73 & 14 13 36.7 & +70 29 50 & RX J1413.6+7029 & 1994/03/06 & McD2.1 & ES2 G22 &
35 & \citet{gru99b} \\
74 & 14 17 59.5 & +25 08 12 & NGC 5548 & 1995/02/25 & McD2.1 & ES2 G4 & 20 & \\
75 & 14 24 03.8 & $-$00 26 58 & QSO 1421$-$0013 & 2002/02/14 & TLS2.0 & NFRS V200 &
4$\times$15 & \\
76 & 14 27 25.0 & +19 49 53 & Mkn 813 & 1999/03/21 & McD2.1 & ES2 G4 & 30 & \\
77 & 14 31 04.1 & +28 17 14 & Mkn 684 & 1995/03\tablenotemark{8} & McD2.1 & ES2
G4 & 150\tablenotemark{8} & \\ 
78 & 14 42 07.5 & +35 26 23 & Mkn 478 & 1995/03\tablenotemark{9} & McD2.1 & ES2
G4 & 150\tablenotemark{9} & \tablenotemark{9} \\
79 & 14 51 08.8 & +27 09 27 & PG 1448+273 & 2002/02/14 & TLS2.0 & NFRS V200 &
2$\times$10 & \\
80 & 15 04 01.2 & +10 26 16 & Mkn 841 & 1999/03/19 & McD2.1 & ES2 G4 & 30 & \\
81 & 15 29 07.5 & +56 16 07 & SBS 1527+564 & 1998/03/04 & McD2.1 & ES2 G4 & 45 &
seeing$\approx~3^{''}$ \\
82 & 15 59 09.7 & +35 01 48 & Mkn 493 & 1997/03/14 & McD2.1 & ES2 G4 & 30 &
clouds \\
83 & 16 13 57.2 & +65 43 11 & Mkn 876 & 1995/04/01 & McD2.1 & ES2 G4 & 25 & \\
84 & 16 18 09.4 & +36 19 58 & RX J1618.1+3619 & 1999/03/22 & McD2.1 & ES2 G4 &
45 & clouds \\
85 & 16 19 51.3 & +40 58 48 & KUG 1618+40 & 1999/03/20 & McD2.1 & ES2 G4 & 50 &
\\
86 & 16 27 56.1 & +55 22 32 & PG 1626+554 & 1996/11/18 & HST & FOS  & 4.25 &
\citet{sha03} \\
87 & 16 29 01.3 & +40 08 00 & EXO 1627+4014 & 1999/03/14 & McD2.7 & LCS  & 30
& \\
88 & 17 02 31.1 & +32 47 20 & RX J1702.5+3247 & 1999/03/21 & McD2.1 & ES2 G4 &
2$\times$30 & \\
89 & 21 32 27.9 & +10 08 20 & II Zw 136 & 1999/09/13 & ESO1.52 & B\&C G23 & 30 &
\\
90 & 21 46 36.0 & $-$30 51 41 & RX J2146.6$-$3051 & 1999/09/14 & ESO1.52 & B\&C
G23 & 30 & light clouds \\
91 & 22 07 45.0 & $-$32 35 01 & ESO 404$-$G029 & 1999/09/13 & ESO1.52 & B\&C G23 &
45 & \\
92 & 22 09 07.6 & $-$27 48 36 & NGC 7214 & 1999/09/14 & ESO1.52 & B\&C G23 & 20 &
light cirrus \\
93 & 22 16 53.2 & $-$44 51 57 & RX J2216.8$-$4451 & 1995/09\tablenotemark{10} & 
ESO1.52 & B\&C G23 & 265\tablenotemark{10} & \\
94 & 22 17 56.6  & $-$59 41 30 & RX J2217.9$-$5941 & 1995/09/21 & ESO1.52 & B\&C
G23 & 40 & \\
95 & 22 30 40.3 & $-$39 42 52 & PKS 2227$-$399 & 1999/09/14 & ESO1.52 & B\&C G23 &
2$\times40$ & light cirrus \\
96 & 22 42 37.7 & $-$38 45 16 & RX J2242.6$-$3845 & 1995/09\tablenotemark{11} &
ESO1.52 & B\&C G23 &  300\tablenotemark{11} & \tablenotemark{11} \\
97 & 22 45 20.3 & $-$46 52 12 & RX J2245.2$-$4652 & 1995/09/28 & ESO1.52 & B\&C
G23 & 30 & light clouds \\
98 & 22 48 41.2 & $-$51 09 53 & RX J2248.6$-$5109 & 1992/10/19 & ESO2.2 & EFOSC
1,8,10 & 5,30,30 & \citet{gru99b} \\
99 & 22 57 39.0 & $-$36 56 07 & MS 2254-36 & 1995/09\tablenotemark{12} & ESO1.52
& B\&C G23 & 85\tablenotemark{12} & \tablenotemark{12} \\
100 & 22 58 45.4 & $-$26 09 14 & RX J2258.7$-$2609 & 1992/08/26 & ESO2.2 & EFOSC
1,8,10 & 5,20,20 & \citet{gru99b} \\
101 & 23 01 36.2 & $-$59 13 20 & RX J2301.6$-$5913 & 1999/06/19 & CTIO4.0 & R-C spec 
& 10 & \\
102 & 23 01 52.0 & $-$55 08 31 & RX J2301.8$-$5508 & 1995/09\tablenotemark{13} &
ESO1.52 & B\&C G23 & 255\tablenotemark{13} & \tablenotemark{13} \\
103 & 23 04 37.3 & $-$35 01 13 & RX J2304.6$-$3501 & 1992/08/25 & ESO2.2 & EFOSC
1,8,10 & 15,30,30 & \citet{gru99b} \\
104 & 23 12 34.8 & $-$34 04 20 & RX J2312.5$-$3404 & 1999/09/13 & ESO1.52 & B\&C
G23 & 2$\times$30 & \\
105 & 23 17 49.9 & $-$44 22 28 & RX J2317.8$-$4422 & 1995/09/19 & ESO1.52 & B\&C
G23 & 50 & \\
106 & 23 25 11.8 & $-$32 36 35 & RX J2325.2$-$3236 & 1993/10/14 & ESO2.2 & EFOSC
1,4 & 5,30 & \citet{gru99b} \\
107 & 23 25 24.2 & $-$38 26 49 & IRAS 23226-3843 & 1999/06/21 & CTIO4.0 & R-C spec
& 10 & clouds \\
108 & 23 43 28.6 & $-$14 55 30 & MS 23409$-$1511 & 1995/09 &
ESO1.52 & B\&C G23 & 3$\times$40 & \\
109 & 23 49 24.1 & $-$31 26 03 & RX J2349.4$-$3126 & 1992/08/21 & ESO2.2 & EFOSC
1,8,10 & 30,30,30 & \citet{gru99b} \\
110 & 23 57 28.0 & $-$30 27 40 & AM 2354$-$304 & 1999/09/12 & ESO1.52 & B\&C G23 &
45 & seeing$\approx~3^{''}$ 
\enddata

\tablenotetext{1}{Mean of four spectra observed for a total of 160 min under
different weather conditions}
\tablenotetext{2}{Mean of three spectra observed for a total of 110 min under
different weather conditions}
\tablenotetext{3}{Mean of six spectra observed for a total of 190 min under
different weather conditions}
\tablenotetext{4}{Mean of five spectra observed for a total of 280 min under
different weather conditions}
\tablenotetext{5}{Mean of six spectra observed for a total of 155 min under
different weather conditions}
\tablenotetext{6}{Mean of seven spectra for a total observing time of 310 min
under different weather conditions.}
\tablenotetext{7}{IDS: Image Dissector Scanner; \citet{wil85}}
\tablenotetext{8}{Mean of five spectra with a total observing time of 150 min.}
\tablenotetext{9}{Mean of six spectra observed for a total of 150 min
under different weather conditions}
\tablenotetext{10}{Mean of six spectra observed for a total of 265 min.}
\tablenotetext{11}{Mean of five spectra observed for 60 min each under different
weather conditions}
\tablenotetext{12}{Mean of three spectra observed for a total of 85 min under
different weather conditions. }
\tablenotetext{13}{Mean of seven spectra observed for a total of 255 min under
different weather conditions.}

\end{deluxetable}

\begin{deluxetable}{rlcrlrlrlrrcc}
\tabletypesize{\scriptsize}
\tablecaption{Line measurements
 of the soft X-ray AGN sample  
 \label{line_sum}}
\tablewidth{0pt}
\tablehead{
& & & \multicolumn{4}{c}{FWHM} & \multicolumn{4}{c}{Equivalent width} & & 
\\
\colhead{\rb{No.}} &   \colhead{\rb{Name}} & \colhead{\rb{z}} 
& \multicolumn{2}{c}{H$\beta$} &
\multicolumn{2}{c}{[OIII]} & \multicolumn{2}{c}{H$\beta$} & \colhead{[OIII]} & \colhead{FeII} 
& \colhead{\rb{$\rm \frac{[OIII]}{H\beta}$}} & 
\colhead{\rb{$\rm \frac{FeII}{H\beta}$}} 
\\
}
\startdata
1 & Mkn 335 & 0.026 & 1710 & \pl140 & 425 & \pl65 & 72 & \pl10 & 25 & 50 & 0.33 & 0.62  \\
2 & ESO 242--G008 & 0.059 & 3000 & \pl200 & 320 & \pl50 & 59 & \pl5 & 52 & 24 & 0.87 & 0.43  \\
3 & Ton S 180 & 0.062 & 970 & \pl100 & 630 & \pl60 & 37 & \pl8 & 4 & 30 & 0.10 & 0.90 \\ 
4 & QSO 0056--36 & 0.165 & 4550 & \pl250 & 360 & \pl50 & 45 & \pl5 & 3 & 20 & 0.06 & 0.49  \\
5 & RX J0100.4--5113 & 0.062 & 3190 & \pl630 & 550 & \pl50 & 36 & \pl10 & 15 & 35 & 0.36 & 0.97 \\
6 & RX J0105.6--1416 & 0.070 & 2600 & \pl300 & 420 & \pl60 & 56 & \pl5 & 70 & 12 & 1.21 & 0.22  \\
7 & RX J0117.5--3826 & 0.225 & 900 & \pl100 & 420 & \pl50 & 50 & \pl10 & 8 & 32 & 0.15 & 0.72  \\
8 & MS 0117--28 & 0.349 & 1681 & \pl260 & 970 & \pl250 & 48 &\pl2 & 4 & 40 & 0.07 & 0.86  \\
9 & RX J0128.1--1848 & 0.046 & 2620 & \pl100 & 400 & \pl60 & 38 & \pl10 & 83 & 33 & 2.18 & 0.89 \\
10 & IRAS F01267--217 & 0.093 & 2520 & \pl160 & 360 & \pl50 & 60 & \pl2 & 23 & 33 & 0.32 & 0.60  \\
11 & RX J0148.3--2758 & 0.121 & 1030 & \pl100 & 700 & \pl500 & 82 & \pl2 & 5 & 51 & 0.05 & 0.69  \\
12 & RX J0152.4--2319 & 0.113 & 2890 & \pl250 & 615 & \pl50 & 57 & \pl10 & 33 & 37 & 0.54 & 0.68 \\
13 & Mkn 1044 & 0.017 & 1310 & \pl100 & 440 & \pl100 & 57 & \pl2 & 7 & 40 & 0.12 & 0.77  \\
14 & Mkn 1048 & 0.042 & 5670 & \pl160 & 435 & \pl60 & 68 & \pl2 & 48 & 15 & 0.63 & 0.23  \\
15 & RX J0311.3--2046 & 0.070 & 4360 & \pl680 & 640 & \pl75 & 40 & \pl2  & 15 & 11 & 0.35 & 0.29   \\
16 & RX J0319.8--2627 & 0.079 & 3100 & \pl1000 & 420 & \pl100 & 48 & \pl4 & 19 & 41 & 0.37 & 0.92  \\
17 & RX J0323.2--4931 & 0.071 & 1680 & \pl250 & 260 & \pl50 & 62 & \pl3 & 16 & 37 & 0.22 & 0.65  \\
18 & ESO 301--G013 & 0.059  & 2410 & \pl680 & 530 & \pl50 & 73 & \pl10 & 121 & 36 & 1.49 & 0.50  \\
19 & VCV 0331--373 & 0.064 & 1880 & \pl110 & 210 & \pl50 & 57 & \pl6 & 22 & 36 & 0.31 & 0.61  \\
20 & RX J0349.1--4711 & 0.299 & 1700 & \pl530 & 700 & \pl50 & 52 & \pl20 & 38 & 67 & 0.43 & 1.16  \\
21 & Fairall 1116 & 0.059 & 4310 & \pl630 & 340 & \pl50 & 67 & \pl5 & 18 & 22 & 0.24 & 0.34  \\
22 & ESO 359--G019 & 0.055 & 9430 & \pl760 & 405 & \pl60 & 78 & \pl10 & 27 & 6 & 0.40 & 0.08  \\
23 & RX J0412.7--4712 & 0.132 & 3520 & \pl940 & 140 & \pl50 & 42 & \pl4 & 41 & 5 & 0.89 & 0.12  \\
24 & 1H 0419--577 & 0.104 & 2580 & \pl200 & 500 & \pl50 & 51 & \pl2 & 82 & 5 & 1.42 & 0.10  \\
25 & Fairall 303 & 0.040 & 1450 & \pl120 & 140 & \pl50 & 100 & \pl10 & 35 & 61 & 0.30 & 0.64  \\
26 & RX J0437.4--4711 & 0.052 & 3990 & \pl500 & 230 & \pl50 & 41 & \pl2 & 11 & 23 & 0.23 & 0.55  \\
27 & RX J0439.6--5311 & 0.243 & 700 & \pl140 & 1380 & \pl360 & 8 & \pl2 & 4 & 21 & 0.47 & 2.66 \\
28 & H 0439--272 & 0.084 & 2550 & \pl150 & 340 & \pl70 & 47 & \pl2 & 88 & 12 & 1.93 & 0.25  \\
29 & 1 ES 0614--584 & 0.057 & 1080 & \pl100 & 200 & \pl50 & 29 & \pl5 & 4 & 27 & 0.13 & 1.00  \\
30 & RX J0859.0+4846 & 0.083 & 2990 & \pl150 & 130 & \pl130 & 74 & \pl3 & 42 & 24 & 0.50 & 0.31 \\
31 & RX J0902.5--0700 & 0.089 & 1860 & \pl150 & 230 & \pl130 & 73 & \pl10 & 21 & 45 & 0.27 & 0.68  \\
32 & Mkn 110 & 0.035 & 1760 & \pl50 & 170 & \pl140 & 65 & \pl7 & 52 & 18 & 0.71 & 0.26  \\
33 & PG 0953+414 & 0.234 & 3000 & \pl220 & 520 & \pl70 & 59 & \pl2 & 14 & 28 & 0.22 & 0.51  \\
34 & RXJ1005.7+4332 & 0.178 & 2740 & \pl250 & 950 & \pl100 & 36 & \pl2 & 2 & 44 & 0.07 & 1.41  \\
35 & RX J1007.1+2203 & 0.083 & 1460 & \pl270 & 170 & \pl130 & 42 & \pl20 & 16 & 30 & 0.27 & 0.69  \\
36 & CBS 126 & 0.079 & 2980 & \pl200 & 460 & \pl70 & 57 & \pl4 & 18 & 16 & 0.29 & 0.30  \\
37 & HS 1019+37 & 0.135 & 6200 & \pl2000 & 800 & \pl100 & 30 & \pl4 & 18 & 11 & 0.64 & 0.35  \\
38 & Mkn 141 & 0.042 & 3600 & \pl110 & 450 & \pl70 & 34 & \pl2 & 15 & 22 & 0.40 & 0.70  \\
39 & Mkn 142 & 0.045 & 1620 & \pl120 & 260 & \pl120 & 48 & \pl1 & 8 & 35 & 0.16 & 0.78  \\
40 & RX J1034.6+3938 & 0.044 & 700 & \pl110 & 340 & \pl100 & 12 &\pl2 & 24 & 18 & 1.65 & 1.37  \\
41 & RX J1117.1+6522 & 0.147 & 1650 & \pl170 & 1000 & \pl150 & 56 & \pl1 & 10 & 48 & 0.18 & 0.99  \\
42 & PG 1115+407 & 0.154 & 1740 & \pl180 & 340 & \pl100 & 53 & \pl8 & 6 & 49 & 0.10 & 0.98   \\
43 & Ton 1388 & 0.177 & 2770 & \pl140 & 950 & \pl110 & 73 &\pl3 & 8 & 46 & 0.10 & 0.72  \\
44 & EXO 1128+69 & 0.045 & 2130 & \pl150 & 200 & \pl140 & 28 & \pl15 & 25 & 15 & 0.84 & 0.55  \\
45 & B2 1128+31 & 0.289 & 3400 & \pl200 & 570 & \pl50 & 27 & \pl10 & 14 & 16 & 0.39 & 0.59 \\
46 & SBS 1136+579 & 0.116 & 2600 & \pl1000 & 400 & \pl100 & 30 & \pl6 & 23 & 25 & 0.70 & 0.86  \\
47 & Z 1136+3412 & 0.033 & 1450 & \pl125 & 190 & \pl130 & 41 & \pl20 & 16 & 43 & 0.30 & 1.02 \\
48 & Was 26 & 0.063 & 2200 & \pl440 & 220 & \pl150 & 37 & \pl5 & 136 & 21 & 3.84 & 0.60  \\
49 & CASG 855 & 0.040 & 4040 & \pl1000 & 200 & \pl140 & 18 & \pl10 & 27 & 13 & 1.25 & 0.61  \\
50 & Mkn 1310 & 0.019 & 3000 & \pl800 & 150 & \pl130 & 37 & \pl3 & 45 & 23 & 1.28 & 0.62 \\
51 & NGC 4051 & 0.002 & 1170 & \pl100 & 370 & \pl125 & 29 & \pl3 & 29 & 27 & 0.92 & 0.94  \\
52 & GQ Com & 0.165 & 3870 & \pl350 & 290 & \pl110 & 38 & \pl6 & 50 & 19 & 1.19 & 0.45  \\
53 & RX J1209.8+3217 & 0.145 & 1320 & \pl110 & 820 & \pl100 & 58 & \pl4 & 34 & 58 & 0.56 & 1.09  \\
54 & PG 1211+143 & 0.082 & 1900 & \pl150 & 280 & \pl130 & 70 & \pl1 & 12 & 32 & 0.14 & 0.50  \\
55 & Mkn 766 & 0.013 & 1100 & \pl200 & 270 & \pl120 & 14 &\pl2 & 52 & 22 & 3.58 & 1.56 \\
56 & 3C 273 & 0.158 & 3050 & \pl200 & 970 & \pl130 & 41 &\pl1  & 5 & 23 & 0.12 & 0.57  \\
57 & RX J1231.6+7044 & 0.208 & 4260 & \pl1250 & 540 & \pl60 & 35 & \pl10 & 70 & 19 & 1.56 & 0.37 \\
58 & Mkn 771 & 0.064 & 3200 & \pl100 & 210 & \pl150 & 73 & \pl2 & 21 & 28 & 0.28 & 0.42 \\
59 & CBS 150 & 0.290 & 1350 & \pl100 & 460 & \pl50 & 47 & \pl3 & 15 & 38 & 0.25 & 0.82  \\
60 & MCG+08--23--067 & 0.030 & 730 & \pl140 & 600 & \pl60 & 9 & \pl3 & 24 & 10 & 2.38 & 1.08   \\
61 & NGC 4593 & 0.009 & 4910 & \pl300 & 180 & \pl150 & 25 & \pl1 & 17 & 11 & 0.66 & 0.43  \\
62 & IRAS F12397+3333 & 0.044 & 1640 & \pl250 & 510 & \pl100 & 26 & \pl10 & 53 & 44 & 1.85 & 1.79 \\
63 & PG 1244+026 & 0.049 & 830 & \pl50 & 330 & \pl90 & 20 & \pl7 & 22 & 27 & 0.88 & 1.28  \\
64 & RX J1304.2+0502 & 0.229 & 1300 & \pl800 & 820 & \pl800 & 36 & \pl10 & 11 & 43 & 0.26 & 1.16  \\
65 & PG 1307+085 & 0.155 & 3860 & \pl220 & 360 & \pl65 & 70 & \pl2 & 30 & 19 & 0.38 & 0.28  \\
66 & RX J1314.3+3429 & 0.075 & 1240 & \pl140 & 260 & \pl120 & 56 & \pl1 & 13 & 39 & 0.21 & 0.78   \\
67 & RX J1319.9+5235 & 0.092 & 950 & \pl100 & 220 & \pl170 & 19 & \pl3 & 35 & 20 & 1.60 & 0.94  \\
68 & PG 1322+659 & 0.168 & 3100 & \pl370 & 130 & \pl130 & 63 & \pl4 & 6 & 25 & 0.10 & 0.43 \\
69 & IRAS 1334+2438 & 0.108 & 2800 & \pl180 & 900 & \pl100 & 47 & \pl3 & 7 & 55 & 0.14 & 1.25   \\
70 & Ton 730 & 0.087 & 3420 & \pl125 & 300 & \pl120 & 52 & \pl4 & 18 & 30 & 0.31 & 0.59  \\
71 & RX J1355.2+5612 & 0.122 & 1100 & \pl100 & 425 & \pl75 & 17 & \pl3 & 95 & 38 & 3.35 & 1.62  \\
72 & PG 1402+261 & 0.164 & 1623 & \pl145 & 710 & \pl100 & 75 & \pl5 & 2 & 72 & 0.02 & 1.10  \\
73 & RX J1413.6+7029 & 0.107 & 4400 & \pl1000 & 175 & \pl50 & 31 & \pl10 & 43 & 32 & 1.49 & 0.97  \\
74 & NGC 5548 & 0.017 & 6460 & \pl200 & 450 & \pl75 & 44 & \pl5 & 43 & 9 & 0.87 & 0.19  \\
75 & QSO 1421--0013 & 0.151 & 1500 & \pl150 & 630 & \pl350 & 69 & \pl3 & 12 & 44 & 0.14 & 0.70  \\
76 & Mkn 813 & 0.111 & 5940 & \pl1400 & 1100 & \pl470 & 38 & \pl5 & 14 & 10 & 0.31 & 0.26  \\
77 & Mkn 684 & 0.046 & 1260 & \pl130 & 170 & \pl140 & 40 & \pl1 & 5 & 52 & 0.12 & 1.50  \\
78 & Mkn 478 & 0.077 & 1630 & \pl150 & 575 & \pl70 & 52 & \pl1 & 9 & 46 & 0.17 & 0.97  \\
79 & PG 1448+273 & 0.065 & 1330 & \pl120 & 260 & \pl200 & 29 & \pl10 & 28 & 25 & 0.83 & 0.94  \\
80 & Mkn 841 & 0.036 & 6000 & \pl1500 & 150 & \pl130 & 64 & \pl20 & 53 & 13 & 0.77 & 0.21  \\
81 & SBS 1527+564 & 0.100 & 2760 & \pl420 & 200 & \pl150 & 70 & \pl5 & 213 & 55 & 3.31 & 0.70  \\
82 & Mkn 493 & 0.032 & 800 & \pl100 & 280 & \pl130 & 17 & \pl5 & 6 & 21 & 0.31 & 1.16  \\
83 & Mkn 876 & 0.129 & 7200 & \pl2000 & 540 & \pl60 & 67 & \pl8 & 17 & 19 & 0.25 & 0.31  \\
84 & RX J1618.1+3619 & 0.034 & 950 & \pl100 & 200 & \pl150 & 10 & \pl5 & 9 & 13 & 0.76 & 1.14  \\
85 & KUG 1618+40 & 0.038 & 1820 & \pl100 & 220 & \pl150 & 48 & \pl5 & 33 & 40 & 0.53 & 0.67  \\
86 & PG 1626+554 & 0.133 & 3460 & \pl1750 & 500 & \pl500 & 62 & \pl20 & 3 & 33 & 0.05 & 0.50  \\
87 & EXO 1627+4014 & 0.272 & 1450 & \pl200 & 200 & \pl140 & 63 & \pl10 & 23 & 31 & 0.35 & 0.49 \\
88 & RX J1702.5+3247 & 0.164 & 1680 & \pl140 & 1110 & \pl100 & 69 & \pl5 & 7 & 63 & 0.08 & 0.98   \\
89 & II Zw 136 & 0.065 & 2210 & \pl250 & 460 & \pl60 & 65 & \pl2 & 16 & 38 & 0.23 & 0.63  \\
90 & RX J2146.6--3051 & 0.075 & 3030 & \pl250 & 400 & \pl60 & 76 & \pl4 & 138 & 28 & 1.75 & 0.35  \\
91 & ESO 404--G029 & 0.063 & 6100 & \pl300 & 280 & \pl80 & 35 & \pl5 & 20 & 14 & 0.55 & 0.41 \\
92 & NGC 7214 & 0.023 & 4700 & \pl250 & 530 & \pl80 & 22 & \pl3 & 13 & 18 & 0.57 & 0.87  \\
93 & RX J2216.8--4451 & 0.136 & 1630 & \pl130 & 700 & \pl50 & 54 & \pl2 & 18 & 58 & 0.29 & 1.13  \\
94 & RX J2217.9--5941 & 0.160 & 1430 & \pl60 & 900 & \pl40 & 37 & \pl5 & 11 & 49 & 0.27 & 0.96  \\
95 & PKS 2227--399 & 0.318 & 3710 & \pl1500 & 380 & \pl50 & 58 & \pl15 & 207 & 23 & 3.60 & 0.39  \\
96 & RX J2242.6--3845 & 0.221 & 1900 & \pl200 & 380 & \pl90 & 69 & \pl2 & 24 & 63 & 0.31 & 1.01  \\
97 & RX J2245.2--4652 & 0.201 & 2250 & \pl300 & 740 & \pl60 & 55 & \pl6 & 28 & 31 & 0.44 & 0.59  \\
98 & RX J2248.6--5109 & 0.102 & 2350 & \pl330 & 260 & \pl30 & 53 & \pl5 & 61 & 18 & 1.07 & 0.34  \\
99 & MS 2254--36 & 0.039 & 1530 & \pl120 & 510 & \pl60 & 47 & \pl3 & 20 & 24 & 0.41 & 0.53  \\
100 & RX J2258.7--2609 & 0.076 & 2030 & \pl180 & 300 & \pl40 & 67 & \pl10 & 95 & 35 & 1.39 & 0.48  \\
101 & RX J2301.6--5913 & 0.149 & 7680 & \pl1500 & 630 & \pl40 & 60 & \pl10 & 20 & 7 & 0.35 & 0.12  \\
102 & RX J2301.8--5508 & 0.140 & 1750 & \pl200 & 280 & \pl70 & 39 & \pl5 & 5 & 54 & 0.12 & 1.55  \\
103 & RX J2304.6--3501 & 0.042 & 1450 & \pl100 & 250 & \pl40 & 31 & \pl5 & 81 & 31 & 2.39 & 0.91  \\
104 & RX J2312.5--3404 & 0.202 & 4200 & \pl950 & 390 & \pl50 & 33 & \pl7 & 29 & 7 & 0.75 & 0.22  \\
105 & RX J2317.8--4422 & 0.132 & 1010 & \pl150 & 330 & \pl60 & 45 & \pl2 & 14 & 50 & 0.28 & 1.09  \\
106 & RX J2325.2--3236 & 0.216 & 3010 & \pl300 & 120 & \pl120 & 86 &\pl15  & 60 & 23 & 0.50 & 0.23  \\
107 & IRAS 23226--3843 & 0.036 & 9500 & \pl4500 & 320 & \pl100 & 37 & \pl10 & 6 & 15 & 0.16 & 0.45  \\
108 & MS23409--1511 & 0.137 & 1030 & \pl100 & 690 & \pl50 & 32 & \pl6 & 7 & 35 & 0.18 & 1.18  \\
109 & RX J2349.4--3126 & 0.135 & 4200 & \pl2000 & 500 & \pl40 & 39 & \pl15 & 27 & 18 & 0.68 & 0.48  \\
110 & AM 2254--304 & 0.033 & 2400 & \pl200 & 250 & \pl100 & 32 & \pl7 & 3 & 24 & 0.09 & 0.81 
\enddata

FWHM are given in \kms~ and the equivalent
widths are given in \AA. 
\end{deluxetable}

\begin{deluxetable}{rlcccccc}
\tabletypesize{\scriptsize}
\tablecaption{X-ray spectral index \ax, luminosities, and X-ray variability parameter $\chi^2/\nu$
 of the soft X-ray AGN sample  
 \label{lum_sum}}
\tablewidth{0pt}
\tablehead{
\colhead{No.} &   \colhead{Name} & \colhead{\ax} 
 & \colhead{log $L_{\rm X}$} & 
\colhead{log $\lambda L_{5100}$} & \colhead{log $L_{\rm bol}$} & 
\colhead{log $L_{\rm H\beta}$} & \colhead{$\chi^2/\nu$}
\\
}
\startdata
1 & Mkn 335 & 2.10 & 36.73 &  36.63 & 38.21 & 34.87 & 2.92 \\
2 & ESO 242--G008 & 1.56 & 36.54 & 36.56 & 37.45 & 34.69 & 2.31 \\
3 & Ton S 180 & 1.89  & 37.13 & 37.27 & 38.71 & 35.22 & 3.47 \\ 
4 & QSO 0056--36 & 1.72  & 37.50 & 37.93 & 39.30 & 35.95  & 1.48 \\
5 & RX J0100.4--5113 & 1.73  & 36.91 & 36.86 & 38.04 & 34.82 & 1.04 \\
6 & RX J0105.6--1416 & 1.29 & 37.11 & 36.86 & 38.48 & 35.02 & 1.24  \\
7 & RX J0117.5--3826 & 2.09 & 37.90 & 37.45 & 39.07 & 35.50 & 1.24  \\
8 & MS 0117--28 & 2.27  & 37.97 & 38.18 & 39.49 & 36.21 & 1.07 \\
9 & RX J0128.1--1848 & 1.55  & 36.37 & 36.20 & 37.09 & 34.14 & 1.04 \\
10 & IRAS F01267--217 & 1.43 & 36.97 & 37.22 & 38.31 & 35.41 & 1.17 \\
11 & RX J0148.3--2758 & 2.12 & 38.20 & 37.45 & 38.71 & 35.71 & 1.36 \\
12 & RX J0152.4--2319 & 1.67  & 37.18 & 37.38 & 38.47 & 35.51 & 1.64 \\
13 & Mkn 1044 & 1.74 & 36.23 & 36.16 &  37.20 & 34.24 & 2.48 \\
14 & Mkn 1048 & 1.53 & 36.71 & 36.80 & 38.25 & 35.04 & 1.92 \\
15 & RX J0311.3--2046 & 1.47 & 36.56 & 36.94 & 38.06 & 34.91 & 0.54   \\
16 & RX J0319.8--2627 & 1.79 & 36.91 & 36.92 & 37.86 & 34.95 & 1.86 \\
17 & RX J0323.2--4931 & 2.03 & 37.00 & 36.58 & 37.62 & 34.72 & 3.13 \\
18 & ESO 301--G013 & 2.01  & 36.76 & 36.75 & 37.75 & 34.97 & 1.24 \\
19 & VCV 0331--373 & 1.59  & 36.60 & 36.53 & 37.68 & 34.69 & 1.10  \\
20 & RX J0349.1--4711 & 2.45  & 37.61 & 37.68 & 39.25 & 35.89 & 2.14  \\
21 & Fairall 1116 & 2.48  & 37.29 & 36.94 & 38.00 & 35.12  & 1.63 \\
22 & ESO 359--G019 & 1.41  & 37.13 & 36.42 & 37.53 & 34.60 & 2.00 \\
23 & RX J0412.7--4712 & 1.66 & 37.24 & 37.46 & 38.23 & 35.45 & 2.24 \\
24 & 1H 0419--577 & 2.20  & 37.83 & 37.87 & 39.38 & 36.00 & 1.33 \\
25 & Fairall 303 & 1.51  & 36.32  & 36.10 & 37.25 & 34.50 & 2.81 \\
26 & RX J0437.4--4711 & 2.09 & 36.60 & 36.84 & 38.01 & 34.83 & 1.19 \\
27 & RX J0439.6--5311 & 2.39 &  37.72 & 37.31 & 38.97 &  34.60 & 1.32 \\
28 & H 0439--272 & 1.51 & 36.91 & 37.12 & 37.92 & 35.13  & 1.48 \\
29 & 1 ES 0614--584 & 2.46  & 36.92 & 36.26 & 37.45 & 34.04 & 3.24 \\
30 & RX J0859.0+4846 & 1.46  & 36.94 & 36.91 & 38.18 & 35.17 & 1.14 \\
31 & RX J0902.5--0700 & 2.17  & 37.14 & 36.34 & 37.86 & 34.59 & 1.15 \\
32 & Mkn 110 & 1.29 & 36.53  & 36.64 & 37.61 & 34.86 & 2.02 \\
33 & PG 0953+414 & 1.65  & 37.78  & 38.14 & 39.39 & 36.29 & 1.05 \\
34 & RXJ1005.7+4332 & 1.81 & 37.37 & 37.50 & 38.86 & 35.39 & 1.16 \\
35 & RX J1007.1+2203 & 1.68 &  36.97 & 36.49 & 38.11 & 34.30 & 2.55 \\
36 & CBS 126 & 1.65  & 37.08 &  37.32 & 38.57 & 35.45 & 1.88 \\
37 & HS 1019+37 & 0.98  & 37.37 & 36.95 & 37.73 & 34.76 & 1.48 \\
38 & Mkn 141 & 1.53 & 36.01 &  36.60 & 37.51 & 34.46 & 2.26 \\
39 & Mkn 142 & 1.88 & 36.57 &  36.51 & 37.56 & 34.53 & 3.18 \\
40 & RX J1034.6+3938 & 2.38 &  36.74 & 36.18 & 37.53 & 33.64 & 1.85 \\
41 & RX J1117.1+6522 & 1.89 & 37.06 & 37.29 & 38.59 & 35.38 & 2.97 \\
42 & PG 1115+407 & 2.05 & 37.29 & 37.51 & 38.83 & 35.60 & 0.97  \\
43 & Ton 1388 & 1.65  & 37.56 &  38.08 & 39.64 & 36.36 & 1.98 \\
44 & EXO 1128+69 & 1.60 & 36.64 & 36.54 & 37.55 & 34.32 & 2.03 \\
45 & B2 1128+31 & 1.43  & 37.94 & 37.72 & 38.80 & 35.65 & 1.18 \\
46 & SBS 1136+579 & 1.54  & 36.92 & 36.60 & 37.80 & 34.46 & 1.07 \\
47 & Z 1136+3412 & 1.81  & 36.20 & 35.96 & 37.12 & 34.00 & 3.52 \\
48 & Was 26 & 1.43  & 37.09 &  36.85 & 38.24 & 34.80 & 1.53 \\
49 & CASG 855 & 1.40  & 36.56 & 36.00 & 37.33 & 33.65 & 1.21 \\
50 & Mkn 1310 & 1.39 &  35.84 &  35.62 & 36.77 & 33.55 & 1.82\\
51 & NGC 4051 & 1.62 &  34.29 &  34.57 & 35.55 & 32.40 & 18.5 \\
52 & GQ Com & 1.18  & 37.42 & 37.40 & 38.26 & 35.44 & 1.54 \\
53 & RX J1209.8+3217 & 3.18  & 36.97 & 36.74 & 37.97 & 34.87 & 0.78 \\
54 & PG 1211+143 & 2.00 & 37.41 & 37.42 & 38.95 & 35.66 & 1.76 \\
55 & Mkn 766 & 1.77  & 36.21 & 36.31 & 37.23 & 33.79  & 8.86 \\
56 & 3C 273 & 1.30  & 38.52 &  38.75 & 39.95 & 36.73 & 0.97 \\
57 & RX J1231.6+7044 & 1.38  & 37.89 & 37.10 & 38.53 & 35.10 & 1.22 \\
58 & Mkn 771 & 1.83  & 36.60 & 36.58 & 37.84 & 34.81 & 1.72 \\
59 & CBS 150 & 2.13  & 37.62 &  37.59 & 39.17 & 35.70 & 1.47 \\
60 & MCG+08--23--067 & 1.38  & 35.88 & 35.81 & 37.27 & 33.11 & 1.34  \\
61 & NGC 4593 & 1.47 & 35.74 & 35.45 & 36.53 & 33.73 & 0.20 \\
62 & IRAS F12397+3333 & 2.02  & 36.39 & 36.36 & 37.64 & 34.12 & 1.93 \\
63 & PG 1244+026 & 1.79 & 36.63 & 36.36 & 37.75 & 34.08 & 1.53 \\
64 & RX J1304.2+0502 & 2.38 & 37.66 & 37.40 & 39.02 & 35.31 & 5.08 \\
65 & PG 1307+085 & 1.58  & 37.41 & 37.52 & 39.08 & 35.78 & 0.87 \\
66 & RX J1314.3+3429 & 1.88  & 36.56 & 36.80 & 37.93 &  34.89 & 3.42  \\
67 & RX J1319.9+5235 & 1.60  & 36.84 & 36.45 & 37.51 & 34.09 & 3.37 \\
68 & PG 1322+659 & 2.07  & 37.51 & 37.71 & 38.81 & 35.87 & 1.52 \\
69 & IRAS 1334+2438 & 1.88  & 37.48 & 37.64 & 38.37 & 35.68 & 1.74  \\
70 & Ton 730 & 1.83  & 36.61 &  36.75 & 37.94 & 34.84 & 2.43 \\
71 & RX J1355.2+5612 & 1.93  & 37.04 & 36.98 & 37.60 & 33.83 & 2.27 \\
72 & PG 1402+261 & 1.83 & 37.36 & 37.72 & 39.17 & 35.99 & 0.71 \\
73 & RX J1413.6+7029 & 1.40 & 37.19 & 36.80 & 37.60 & 34.63 & 1.70 \\
74 & NGC 5548 & 1.33 & 36.40 &  36.56 & 37.53 & 34.61 & 1.27 \\
75 & QSO 1421--0013 & 1.72 & 37.48 & 37.34 & 38.76 & 35.56 & 2.34 \\
76 & Mkn 813 & 1.64  & 37.30 & 37.28 & 38.95 & 35.28 & 1.41 \\
77 & Mkn 684 & 1.36 & 36.23 & 36.87 & 38.35 & 34.81 & 1.75 \\
78 & Mkn 478 & 2.08 & 37.45 & 37.46 & 38.88 & 35.53 & 3.71 \\
79 & PG 1448+273 & 1.52 &  36.87 & 37.10 & 38.56 & 35.00 & 1.60 \\
80 & Mkn 841 & 1.50 & 36.28 & 36.64 & 37.79 & 34.85 & 1.00 \\
81 & SBS 1527+564 & 1.46 &  37.01  & 36.41 & 37.86 & 34.69 & 1.34 \\
82 & Mkn 493 & 1.57 &  36.03 &  36.74 & 37.63 & 34.33 & 2.03 \\
83 & Mkn 876 & 1.67 & 37.52 &  37.76 & 38.85 & 35.94  & 1.75 \\
84 & RX J1618.1+3619 & 1.54 & 36.12  & 35.71 & 36.79 & 33.09 & 2.49 \\
85 & KUG 1618+40 & 1.52 & 35.96 & 35.81 & 37.16 & 33.68  & 1.87 \\
86 & PG 1626+554 & 1.79 & 37.26 &  37.54 & 38.88 & 35.74  & 1.17 \\
87 & EXO 1627+4014 & 2.25 & 37.61 & 37.41 & 38.90 & 35.56  & 1.17 \\
88 & RX J1702.5+3247 & 2.13  & 37.47 & 37.29 & 39.22 & 35.51 & 1.29  \\
89 & II Zw 136 & 2.10 & 37.34 & 37.15 & 38.86 & 35.35  & 1.39 \\
90 & RX J2146.6--3051 & 1.59  & 36.92 & 36.72 & 37.61 & 35.00 & 1.62  \\
91 & ESO 404--G029 & 1.37  & 36.77 & 36.77 & 37.52 & 34.64  & 1.78 \\
92 & NGC 7214 & 1.34 & 35.81 & 36.30 & 37.22 & 33.96 & 1.84 \\
93 & RX J2216.8--4451 & 2.48  & 37.59 & 37.17 & 38.55 & 35.27 & 1.76 \\
94 & RX J2217.9--5941 & 2.69  & 37.58 & 37.14 & 38.60 & 35.06 & 4.20 \\
95 & PKS 2227--399 & 1.00  & 37.92 & 37.24 & 38.43 & 35.39 & 1.14 \\
96 & RX J2242.6--3845 & 2.19  & 37.42 & 37.03 & 38.50 & 35.23 & 0.89 \\
97 & RX J2245.2--4652 & 2.55  & 37.88 & 37.79 & 39.02 & 35.94  & 1.52 \\
98 & RX J2248.6--5109 & 1.95  & 37.47 & 37.35 & 38.33 & 35.46 & 1.23 \\
99 & MS 2254--36 & 1.78 & 36.51 & 36.34 & 37.27 & 34.33 & 2.34 \\
100 & RX J2258.7--2609 & 1.50  & 36.83 & 36.74 & 37.62 & 34.97 & 0.52 \\
101 & RX J2301.6--5913 & 1.65  & 37.69 & 37.35 & 38.30 & 35.49 & 1.03 \\
102 & RX J2301.8--5508 & 2.09  & 37.33 & 37.38 & 38.47 & 35.33 & 2.27 \\
103 & RX J2304.6--3501 & 1.65  & 36.16 & 36.08 & 37.08 & 33.98 & 1.94 \\
104 & RX J2312.5--3404 & 1.34  & 37.48 & 37.51 & 39.20 & 35.47 & 2.70 \\
105 & RX J2317.8--4422 & 2.87  & 37.17 & 36.77 & 38.89 & 34.83 & 0.79  \\
106 & RX J2325.2--3236 & 1.92  & 37.47 & 37.36 & 38.78 & 35.75 & 0.27 \\
107 & IRAS 23226--3843 & 1.20  & 36.43 & 36.43 & 36.88 & 34.27 & 0.93 \\
108 & MS23409--1511 & 2.03 & 37.39 & 37.42 & 38.52 & 35.28 & 1.66 \\
109 & RX J2349.4--3126 & 1.67  & 37.11 & 37.17 & 38.00 & 35.09 & 0.80 \\
110 & AM 2254--304 & 1.30 & 35.95 & 36.18 & 36.87 & 34.00 & 1.62
\enddata

All luminosities are given in units of Watts.
\end{deluxetable}

\begin{deluxetable}{lccccccccc}
\tabletypesize{\scriptsize}
\tablecaption{Mean, Standard deviation, and median of the whole sample (110 sources), 
NLS1s (51), and BLS1s (59).
 \label{distr_sum}}
\tablewidth{0pt}
\tablehead{
& \multicolumn{3}{c}{all sources (110)} & \multicolumn{3}{c}{NLS1s (51)} & 
\multicolumn{3}{c}{BLS1s1 (59)}  \\
\colhead{\rb{Property}} & \colhead{Mean} & \colhead{$\sigma$} & 
\colhead{Median} & \colhead{Mean} & \colhead{$\sigma$} & \colhead{Median} & 
\colhead{Mean} & \colhead{$\sigma$} & \colhead{Median} 
}
\startdata
FWHM(H$\beta$) & 2695 & 1760 & 2210 & 1380 & 350 & 1450 & 3830 & 1700 & 3100 \\
FWHM([OIII]) & 431 & 255 & 370 & 445 & 280 & 345 & 420 & 240 & 390 \\
EW(H$\beta$) & 48 & 19 & 47 & 45 & 21 & 47 & 50 & 17 & 47  \\
EW([OIII]) & 32 & 37 & 20 & 24 & 32 & 16 & 38 & 40 & 23 \\
EW(FeII) & 30 & 15 & 28 & 38 & 15 & 38 & 23 & 12 & 22 \\
$\rm [OIII]/H\beta$ & 0.71 & 0.82 & 0.36 & 0.66 & 0.91 & 0.29 & 0.75 & 0.74 & 0.50  \\
FeII/H$\beta$ & 0.73 & 0.42 & 0.65 & 0.99 & 0.40 & 0.96 & 0.50 & 0.28 & 0.48 \\
\ax & 1.77 & 0.39 & 1.67 & 1.96 & 0.41 & 1.93 & 1.62 & 0.30 & 1.56 \\
log $L_{\rm X}$ & 36.99 & 0.64 & 37.04 & 36.94 & 0.71 & 37.00 & 37.03 & 0.57 &
37.09 \\
log $\lambda~L_{5100}$ & 36.93 & 0.64 & 36.91 & 36.81 & 0.66 & 36.80 &
37.03 & 0.61 & 36.94 \\
log $L_{\rm bol}$ & 38.13 & 0.78 & 38.06 & 38.10 & 0.82 & 38.11 & 38.15 & 0.74 &
38.06 \\
log $L_{\rm H\beta}$ & 34.92 & 0.81 & 34.97 & 34.76 & 0.81 & 34.86 & 35.05 & 
0.79 & 35.04  \\
log $L_{\rm [OIII]}$ & 34.53 & 0.73 & 34.54 & 34.29 & 0.63 & 34.36 & 
34.74 & 0.75 & 34.84 \\
log $L_{\rm FeII}$ & 34.70 & 0.79 & 34.66 & 34.72 & 0.79 & 34.87 & 34.68 & 0.79
& 34.65  \\
z & 0.107 & 0.075 & 0.083 & 0.113 & 0.088 & 0.082 & 0.101 & 0.061 & 0.084 \\
$\chi^2/\nu$ & 1.95 & 1.92 & 1.60 & 2.57 & 2.65 & 1.94 & 1.43 & 0.50 & 1.41 \\
\enddata

The FWHM are given in units of \kms, the rest frame
equivalent widths in \AA, and the luminosities
are given in units of Watts.

\end{deluxetable}

\begin{figure*}

\caption{\label{opt_spectra} Optical spectra of the soft X-ray selected sample
which have not been published yet (see Table \ref{obs_sum}). The wavelength is given in
units of \AA~and the flux density is given in units of 
$10^{-19}$ W m$^{-2}$ \AA$^{-1}$. All spectra are shown in the observed frame.
{\bf Please note:} this is a special version for astro-ph that does not contain
the optical and FeII subtracted spectra. The complete paper including the
spectra can be retrievd from
http://www.astronomy.ohio-state.edu/$\sim$dgrupe/research/sample\_paper1.html
}

\end{figure*}

\begin{figure*}

\caption{\label{fe2_spectra} FeII-subtracted spectra, the wavelength is given in
units of \AA~and the flux density is given in units of 
$10^{-19}$ W m$^{-2}$ \AA$^{-1}$.
The dotted line marks the zero line of the FeII template.
{\bf Please note:} this is a special version for astro-ph that does not contain
the optical and FeII subtracted spectra. The complete paper including the
spectra can be retrievd from
http://www.astronomy.ohio-state.edu/$\sim$dgrupe/research/sample\_paper1.html
}
\end{figure*}

\end{document}

%% file: DGrupe_clipfig.tex
\def\clipfig#1{\def\lbracket{[}\def\testit{#1}%
    \ifx\testit\lbracket\let\next=\optclipfig\else\let\next=\stdclipfig\fi%
    \next{#1}}
%
\newcommand {\hclipfig} [7] {\clipfig[#7]{#1}{#2}{#3}{#4}{#5}{#6}}
%
\def\usemodepsfig {\global\def\cfmode{x}\typeout{*** set clipfig to PSFIG mode ***}}
\def\usemodeepsf  {\global\def\cfmode{}\typeout{*** set clipfig to EPSF mode ***}}
\def\useunitmm    {\global\def\cfunit{x}\typeout{*** set clipfig to use mm as unit ***}}
\def\useunitcm    {\global\def\cfunit{}\typeout{*** set clipfig to use cm as unit ***}}
\def\clipfigsettings {\ifx\cfmode\empty\def\ccfmode{EPSF }\else\def\ccfmode{PSFIG }\fi%
    \ifx\cfunit\empty\def\ccfunit{cm }\else\def\ccfunit{mm }\fi%
    \typeout{*** current clipfig settings: \ccfmode mode, using \ccfunit as unit ***}}
%
%
%
%
\def\stdclipfig#1#2#3#4#5#6{\ifx\cfmode\empty%
    \let\next=\eclipfig\else\let\next=\pclipfig\fi%
    \next{#1}{#2}{#3}{#4}{#5}{#6}}
\def\optclipfig#1#2]#3#4#5#6#7#8{\ifx\cfmode\empty%
    \let\next=\ehclipfig\else\let\next=\phclipfig\fi%
    \next{#3}{#4}{#5}{#6}{#7}{#8}{#2}}
%
%
%
\newcommand {\pclipfig}[6] {\ifx\cfunit\empty%
        \psfig{figure=#1.ps,width=#2cm,bbllx=#3cm,bblly=#4cm,bburx=#5cm,%
           bbury=#6cm,clip=}\else%
        \psfig{figure=#1.ps,width=#2mm,bbllx=#3mm,bblly=#4mm,bburx=#5mm,%
           bbury=#6mm,clip=}\fi}
\newcommand {\phclipfig}[7] {\ifx\cfunit\empty%
        \hspace{#7cm}\psfig{figure=#1.ps,width=#2cm,bbllx=#3cm,bblly=#4cm,%
           bburx=#5cm,bbury=#6cm,clip=}\else%
        \hspace{#7mm}\psfig{figure=#1.ps,width=#2mm,bbllx=#3mm,bblly=#4mm,%
           bburx=#5mm,bbury=#6mm,clip=}\fi}
%
%
%
\newcommand {\eclipfig}[6]{%
  \ifx\cfunit\empty\epsfxsize=#2cm\else\epsfxsize=#2mm\fi%
  \epsfclipon\epsfverbosetrue%
  \cfcmtopspts{#3}\cfllxi=\cftempi\cfllxf=\cftempf%
  \cfcmtopspts{#4}\cfllyi=\cftempi\cfllyf=\cftempf%
  \cfcmtopspts{#5}\cfurxi=\cftempi\cfurxf=\cftempf%
  \cfcmtopspts{#6}\cfuryi=\cftempi\cfuryf=\cftempf%
  \def\cfstra{\number\cfllxi.\number\cfllxf}%
  \def\cfstrb{\number\cfllyi.\number\cfllyf}%
  \def\cfstrc{\number\cfurxi.\number\cfurxf}%
  \def\cfstrd{\number\cfuryi.\number\cfuryf}%
  \hbox{\epsfbox[{\cfstra} {\cfstrb} {\cfstrc} {\cfstrd}]{#1.ps}}}
\newcommand {\ehclipfig}[7]{%
  \ifx\cfunit\empty\epsfxsize=#2cm\else\epsfxsize=#2mm\fi%
  \epsfclipon\epsfverbosetrue%
  \cfcmtopspts{#3}\cfllxi=\cftempi\cfllxf=\cftempf%
  \cfcmtopspts{#4}\cfllyi=\cftempi\cfllyf=\cftempf%
  \cfcmtopspts{#5}\cfurxi=\cftempi\cfurxf=\cftempf%
  \cfcmtopspts{#6}\cfuryi=\cftempi\cfuryf=\cftempf%
  \def\cfstra{\number\cfllxi.\number\cfllxf}%
  \def\cfstrb{\number\cfllyi.\number\cfllyf}%
  \def\cfstrc{\number\cfurxi.\number\cfurxf}%
  \def\cfstrd{\number\cfuryi.\number\cfuryf}%
  \ifx\cfunit\empty\hspace{#7cm}\else\hspace{#7mm}\fi%
  \hbox{\epsfbox[{\cfstra} {\cfstrb} {\cfstrc} {\cfstrd}]{#1.ps}}%
  \vspace{-1mm}}
%
%
%
\newdimen\cfllxi \newdimen\cfllyi  \newdimen\cfurxi  \newdimen\cfuryi
\newdimen\cfllxf \newdimen\cfllyf  \newdimen\cfurxf  \newdimen\cfuryf
\newdimen\cftemp \newdimen\cftempi \newdimen\cftempf
\newdimen\cfpspoint \cfpspoint=1bp
%
%
%
\newcommand{\cfcmtopspts}[1]{\ifx\cfunit\empty%
  \cftemp=#1cm\else\cftemp=#1mm\fi%
  \multiply\cftemp10\divide\cftemp\cfpspoint%
  \cftempf=\cftemp\divide\cftemp10\cftempi=\cftemp\multiply\cftemp10%
  \advance\cftempf-\cftemp}
%
%
\def\cfmode{}\def\cfunit{}\clipfigsettings
%